\documentclass[aps,prl,twocolumn,superscriptaddress,floatfix]{revtex4-1}

\usepackage{graphicx}
\usepackage{dcolumn}
\usepackage{bm}
\usepackage{amssymb}
\usepackage{microtype}
\usepackage{xfrac}

\newcommand{\ket}[1]{\ensuremath{|#1\rangle}}

\newcommand{\angstrom}{\text{\normalfont\AA}}
\usepackage{mathtools}
\DeclarePairedDelimiter{\evdel}{\langle}{\rangle}
\newcommand{\ev}{\evdel}

\begin{document}

\title{Effect of Chemical Pressure on the Crystal Electric Field States \\ of Erbium Pyrochlore Magnets}

\author{J.~Gaudet}
\affiliation{Department of Physics and Astronomy, McMaster University, Hamilton, ON, L8S 4M1, Canada}

\author{A.~M.~Hallas}
\affiliation{Department of Physics and Astronomy, McMaster University, Hamilton, ON, L8S 4M1, Canada}

\author{A.~I.~Kolesnikov}
\affiliation{Chemical and Engineering Materials Division, Oak Ridge National Laboratory, Oak Ridge, TN, 37831, USA}

\author{B.~D.~Gaulin}
\affiliation{Department of Physics and Astronomy, McMaster University, Hamilton, ON, L8S 4M1, Canada}
\affiliation{Canadian Institute for Advanced Research, 180 Dundas St. W., Toronto, ON, M5G 1Z7, Canada}
\affiliation{Brockhouse Institute for Materials Research, Hamilton, ON L8S 4M1 Canada}

\date{\today}

\begin{abstract}
We have carried out a systematic study of the crystal electric field excitations in the family of cubic pyrochlores Er$_2B_2$O$_7$, with $B=$~Ge, Ti, Pt, and Sn, using neutron spectroscopy. All members of this family are magnetic insulators based on 4$f^{11}$ Er$^{3+}$ and non-magnetic $B^{4+}$. At sufficiently low temperatures, long-range antiferromagnetic order is observed in each of these Er$_2B_2$O$_7$ pyrochlores. Our inelastic neutron scattering measurements probe the transitions from the ground state doublet to excited crystal electric field states belonging to the $J=15/2$ Hund's rules manifold. This allows us to quantitatively determine the energy eigenvalues and eigenfunctions of these $(2J+1)=16$ states across the Er$_2B_2$O$_7$ series. The different ionic sizes associated with different non-magnetic $B^{4+}$ cations correspond to positive or negative chemical pressure, depending on the relative contraction or expansion of the crystal lattice, which gives rise to different local environments at the Er$^{3+}$ site. Our results show that the $g$-tensor components are XY-like for all four members of the Er$_2B_2$O$_7$ series. However, the XY anisotropy is much stronger for Er$_2$Pt$_2$O$_7$ and Er$_2$Sn$_2$O$_7$ ($\sfrac{g_{\perp}}{g_z} > 25$), than for Er$_2$Ge$_2$O$_7$ and Er$_2$Ti$_2$O$_7$ ($\sfrac{g_{\perp}}{g_z} < 4$). The variation in the nature of the XY-anisotropy in these systems correlates strongly with their ground states, as Er$_2$Ge$_2$O$_7$ and Er$_2$Ti$_2$O$_7$ order into $\Gamma_5$ magnetic structures, while Er$_2$Pt$_2$O$_7$ and Er$_2$Sn$_2$O$_7$ order in the $\Gamma_7$ Palmer-Chalker structure.
\end{abstract}

\maketitle
\section{Introduction}

The cubic pyrochlore lattice is adopted by many magnetic materials with chemical composition $A_2B_2$O$_7$. This lattice is prone to magnetic frustration due to its architecture that consists of two inter-penetrating networks of corner-sharing tetrahedra on which both the $A^{3+}$ and $B^{4+}$ ions independently reside~\cite{Greedan}. Much attention has been devoted to pyrochlore magnets whose spin anisotropy has a local Ising character, such that the moments are constrained to point along the local $\ev{111}$ axis, into or out of the tetrahedra.  When combined with an effective ferromagnetic near-neighbor coupling, such local Ising anisotropy gives rise to spin configurations in which two spins point in and two spins point out of each tetrahedra~\cite{harris1997geometrical}. This can lead to a macroscopic degeneracy analogous to that of proton disorder in the frozen state of water, hence this magnetic state has been dubbed spin ice~\cite{ramirez1999zero,bramwell2001spin}. The scientific interest in spin ice materials, such as Ho$_2$Ti$_2$O$_7$ and Dy$_2$Ti$_2$O$_7$, originates from the observation of classical spin liquid behavior with magnetic excitations that can be mapped onto diffusive magnetic monopoles~\cite{Castelnovo2008,morris2009dirac}.\

Recently, growing interest has been devoted to magnetic pyrochlores with the complementary form of anisotropy, local XY anisotropy where the spins are constrained to point in a plane perpendicular to the local Ising axis~\cite{RevXYHallas}. XY pyrochlore magnets, such as Yb$_2$Ti$_2$O$_7$ and Er$_2$Ti$_2$O$_7$, possess an effective $S=1/2$ degree of freedom~\cite{Ross2011,Savary2012}. The XY pyrochlores have generated much interest for their potential to stabilize a quantum analogue to the spin ice state~\cite{Ross2011,gingras2014quantum}. Furthermore, the XY pyrochlores may harbor a realization of ground state selection by quantum order-by-disorder~\cite{Champion2003,Savary2012,Zhitomirsky2012}.

Underpinning the remarkable physics of the XY pyrochlores is the highly anisotropic nature of their exchange interactions. Classical calculations using an anisotropic exchange Hamiltonian have shown that several different ordered magnetic ground states exist within a narrow subspace of exchange parameters that are relevant to known XY pyrochlore magnets. This has led to theoretical proposals that competition between phases could account for many of their observed magnetic properties~\cite{Jaubert2015,Yan2017}. Experimentally, evidence for phase competition is indeed found within the Er$_2B_2$O$_7$ family of XY pyrochlores. Both Er$_2$Ti$_2$O$_7$ and Er$_2$Ge$_2$O$_7$ order into the $k=0$ $\Gamma_5$ manifold~\cite{Champion2003,Dun2015}, but in a pure $\psi_2$ state for the former~\cite{Poole2007} and likely a pure $\psi_3$ state for the latter~\cite{Dun2015}. Meanwhile, both Er$_2$Pt$_2$O$_7$ and Er$_2$Sn$_2$O$_7$ are known to order into a $k=0$ $\Gamma_7$ manifold~\cite{Hallas2017EPO,Petit2017}. The magnetic structures observed within this family, where only the non-magnetic $B$-site ion is varied, are all ones that exist in the computed anisotropic exchange phase diagram~\cite{Yan2017}. It is therefore interesting to uncover the microscopic origin of the changing magnetic ground states across this family that results from changing the non-magnetic $B$-site cation. To carry out such a detailed study, a key starting point is the precise determination of the spin anisotropy that originates from single-ion physics.

In this paper, we study the single ion properties of the four erbium pyrochlores within this family, Er$_2$Ge$_2$O$_7$, Er$_2$Ti$_2$O$_7$, Er$_2$Pt$_2$O$_7$ and Er$_2$Sn$_2$O$_7$. We measure their crystal electric field excitations using neutron spectroscopic techniques. The analysis of these measurements allows us to determine the full CEF Hamiltonian for each member of this family. We show that the Er$^{3+}$ moments in all four members possess XY-like anisotropy, but the strength of this anisotropy varies significantly across the family. We then show that the degree of XY anisotropy of the Er$^{3+}$ moments correlates strongly with the precise magnetic ground state that is selected.\\

\section{Crystal-field calculation}
The magnetism in Er$_2B_2$O$_7$ originates from the Er$^{3+}$ ions that possess a [Xe]4$f^{11}$ electronic configuration. Using Hund's rules, the total angular momentum of the spin-orbit ground state is $J=\frac{15}{2}$ with $L=6$ and $S=\frac{3}{2}$. This spin-orbit ground state is $2J+1=16$ fold degenerate in the absence of the crystal electric field. The neighboring ions, primarily the O$^{2-}$ that surround the Er$^{3+}$ ion generate the CEF. The form of the CEF Hamiltonian depends on the point group symmetry at the Er$^{3+}$ site, which has two-fold and three-fold rotation axes as well as an inversion symmetry with respect to the local $\ev{111}$ axis. These symmetry operations result in a $D_{3d}$ point-group symmetry. Using the Steven's operator formalism, the CEF Hamiltonian for the $D_{3d}$ point-group symmetry can be written as follows~\cite{Hutchings,Prather,Freeman1962,Walter1984}:

\begin{eqnarray}
\mathcal{H}_{CEF}= A^0_2\alpha_J\ev{r^2}\hat{O}^0_2 + A^0_4\beta_J\ev{r^4}\hat{O}^0_4 +  
\nonumber
\\
A^3_4\beta_J\ev{r^4}\hat{O}^3_4 + A^0_6\gamma_J\ev{r^6}\hat{O}^0_6 +
\nonumber
\\
A^3_6\gamma_J\ev{r^6}\hat{O}^3_6 + A^6_6\gamma_J\ev{r^6}\hat{O}^6_6.
\label{eq: HCEF}
\end{eqnarray}

The $\hat{O}^m_n$ are Steven's operators that are written in terms of $J_+$, $J_-$, and $J_z$ operators~\cite{Stevens}. The $\alpha_J$, $\beta_J$ and $\gamma_J$ are reduced matrix elements that have been previously calculated in Ref.~\cite{Stevens}. The values for $\ev{r^n}$ are listed in Ref.~\cite{freeman1979dirac} and consist of the expected value of the $n$th power of distance between a nucleus and the 4$f$ electron shell. The 16 states of the spin-orbit ground state are split by the CEF into 8 doublets, and these are protected from further degeneracy-breaking by Kramer's theorem. The CEF energy eigenvalues and eigenfunctions associated with these 8 doublets are controlled by the crystal field parameters $A^m_n$, and these can be directly probed with inelastic neutron spectroscopy. The partial differential cross-section for magnetic neutron scattering can be written as follows~\cite{Squires}:
\begin{eqnarray}
\frac{d^2\sigma}{d\Omega dE'} = C\frac{k_f}{k_i}F^2(|Q|)S(|Q|,\hbar \omega),
\end{eqnarray} 
where $\Omega$ is the scattered solid angle, $E'$ is the final neutron energy, $\frac{k_f}{k_i}$ is the ratio of the scattered and incident momentum of the neutron, $C$ is a constant, and $F(|Q|)$ is the magnetic form factor. The scattering function $S(|Q|,\hbar\omega)$ for a CEF transition is given by:
\begin {eqnarray}
S(|Q|,\hbar\omega) = \sum_{i,i'}\frac{\sum_{\alpha}  |\langle i {| J_{\alpha} | i'\rangle |}^2 \mathrm{e}^{-\beta E_{i}} }{\sum_j \mathrm{e}^{-\beta E_{j}}} L(\Delta E + \hbar \omega)
\end{eqnarray} 
where $\alpha = x,y,z$ and $L(\Delta E + \hbar \omega) = L(E_{i} - E_{i'} + \hbar \omega)$ is a Lorentzian function that ensures energy conservation as the neutron induces transitions between the CEF levels $i \rightarrow i'$, which possess a finite energy width or lifetime. \\

In order to determine the CEF parameters, $A^m_n$, we start with a particular set of $A^m_n$ and diagonalize the CEF Hamiltonian to obtain a set of initial energy eigenvalues and eigenfunctions. In this work, we used the $A^m_n$ obtained for Er$_2$Ti$_2$O$_7$ from Ref.~\cite{bertin2012crystal} as our starting CEF parameters. A least squares refinement of the $A^m_n$ values is then performed to minimize the difference between the calculated and observed CEF spectra. This refinement considers both the energies of the transitions as well as their relative intensities. \\

\begin{figure}[tbp]
\linespread{1}
\par
\includegraphics[width=3.2in]{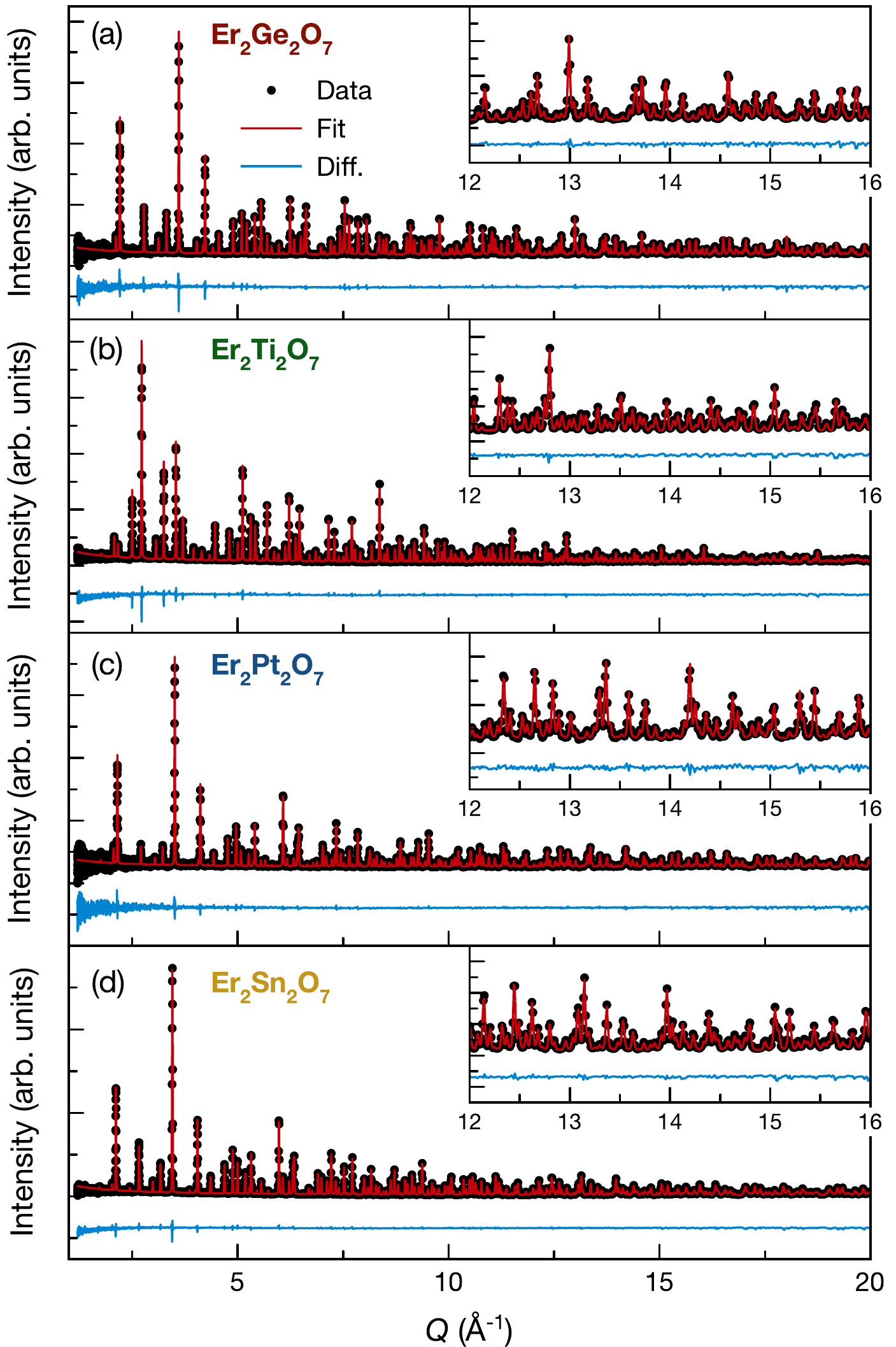}
\par
\caption{Rietveld refinements of the time-of-flight neutron powder diffraction patterns for each of the erbium pyrochlore magnets, Er$_2B_2$O$_7$, $B=$~(a) Ge (b) Ti (c) Pt and (d) Sn. The data sets were collected at 10~K using Bank 2 of the POWGEN diffractometer ($\lambda_{\text{center}} = 1.066$~\AA). The data were refined against the $Fd\bar{3}m$ space group, and the goodness-of-fit parameters for each sample are given in Table~\ref{RietveldTable}. The insets show an expanded view of a high $Q$ region, from 12 to 16~\AA$^{-1}$.}
\label{Refinement}
\end{figure}

\begin{table*}[tbp]
\caption{Summary of fitted structural parameters for the erbium pyrochlore magnets, as determined by Rietveld refinement of their powder neutron diffraction patterns measured at 10 K. The variance in the cubic lattice parameter, $a$, is largely controlled by the radius of the $B$-site cation. The only adjustable atomic coordinate within the pyrochlore ($Fd\bar{3}m$) structure is the $x$ position of the apical oxygens (O1). The four entries of the final columns give the goodness-of-fit parameters for the Rietveld refinements, which are shown in Figure~\ref{Refinement}.}
\label{RietveldTable}
\begin{tabular}{l>{\centering}p{1.5cm}>{\centering}p{1.4cm}p{1.4cm}<{\centering}p{1.6cm}<{\centering}p{1.6cm}<{\centering}p{1.2cm}<{\centering}p{1.2cm}<{\centering}p{1.2cm}<{\centering}p{1.2cm}<{\centering}p{1.2cm}<{\centering}}
\hline
\hline
                  & $B$ (\AA) & \textit{a} (\AA) & O1 $x$ & Er-O1 (\AA) & Er-O2 (\AA) & $\frac{\text{Er-O2}}{\text{Er-O1}}$  &  R$_p$ & R$_{wp}$  & R$_{exp}$ & $\chi^2$ \\ \hline
Er$_2$Ge$_2$O$_7$ & 0.67   & 9.87(5)  & 0.327 & 2.442 & 2.136 & 0.875  & 5.4\% & 4.0\%  & 1.6\%  & 6.5     \\
Er$_2$Ti$_2$O$_7$ & 0.745 & 10.05(3) & 0.330 & 2.463 & 2.176 & 0.883  & 7.7\% & 6.8\%  & 1.7\%  & 15.8     \\
Er$_2$Pt$_2$O$_7$ & 0.765 & 10.13(2) & 0.340 & 2.416 & 2.193 & 0.908  & 4.6\% & 2.8\%  & 1.6\%  & 3.0     \\
Er$_2$Sn$_2$O$_7$ & 0.83  & 10.30(1) & 0.338 & 2.470 & 2.230 & 0.903  & 8.0\% & 7.2\%  & 1.9\%  & 14.6     \\ \hline
\hline
\end{tabular}
\end{table*}

\section{Experimental Details}
The four erbium pyrochlore magnets were all studied in powder form. Large 10~g samples of both Er$_2$Ti$_2$O$_7$ and Er$_2$Sn$_2$O$_7$ were prepared via conventional solid state synthesis~\cite{subramanian1983oxide}. Er$_2$Ge$_2$O$_7$ and Er$_2$Pt$_2$O$_7$ were prepared using a belt-type high pressure apparatus, yielding 2.8~g and 1.1~g samples, respectively. The details of the high pressure synthetic method employed are given in Ref.~\cite{hallas2016xy} for the Er$_2$Ge$_2$O$_7$ sample and in Ref.~\cite{hallas2016relief} for the Er$_2$Pt$_2$O$_7$ sample.\\

Powder neutron diffraction was carried out on each of the four samples using the time-of-flight diffractometer POWGEN at the Spallation Neutron Source at Oak Ridge National Laboratory~\cite{huq2011powgen}. The samples were sealed in vanadium cans in the presence of a helium atmosphere and were each measured for one hour at 10~K. These measurements were performed with a median wavelength of $\lambda = 1.066$~\AA. This configuration allows for momentum transfers between 1.2~\AA$^{-1} < Q < 22.8$~\AA$^{-1}$. Rietveld refinements of the data were carried out using the FullProf~\cite{rodriguez1993recent} software suite.\\ 

The inelastic neutron scattering measurements were performed using the direct geometry, time-of-flight chopper spectrometer, SEQUOIA, at the Spallation Neutron Source~\cite{granroth2010sequoia}. Each sample was measured at two temperatures (5~K and 100~K) and three incident neutron energies (25~meV, 90~meV and 150~meV). For all incident neutron energies, the fine Fermi chopper was used to achieve a maximum energy resolution of $1-2\%$ of initial energy. The samples were packed in aluminum cans in annular geometry and sealed with indium in the presence of a helium atmosphere. An identical empty can was also measured and this served as a background, where all the data presented have had the empty can data set subtracted from them. The inelastic neutron scattering data were reduced using Mantid~\cite{Mantid} and were analyzed using DAVE~\cite{Dave}.\\ 

\section{Results and analysis}

\subsection{Structural refinement by powder neutron diffraction}

\begin{figure*}[htbp]
\linespread{1}
\par
\includegraphics[width=7in]{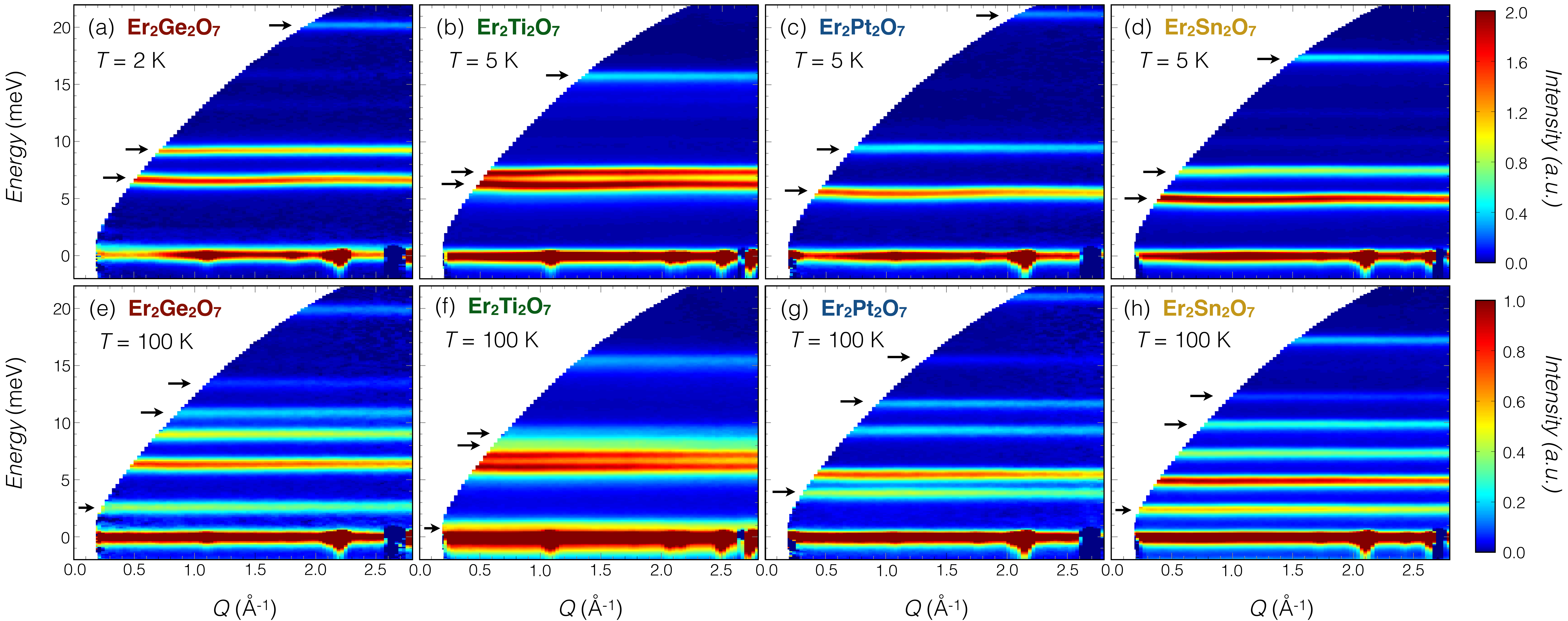}
\par
\caption{Inelastic neutron scattering measurements with an incident energy of $E_i = 25$~meV for the Er$_2B_2$O$_7$ pyrochlores, $B=$~(a,e) Ge, (b,f) Ti, (c,g) Pt, and (d,h) Sn. The top row shows the spectra at base temperature (2~K or 5~K) for each sample. In each case, three ground state crystal electric field excitations are identified at the positions indicated by the arrows. The bottom row show the spectra at 100~K, which is high enough to thermally populate transitions from the first and second excited state levels. The excited state transitions at 100~K are indicated by the arrows. An empty can measurement has been subtracted from all data sets.}
\label{25meV_Slices}
\end{figure*}

\begin{figure*}[htbp]
\linespread{1}
\par
\includegraphics[width=7in]{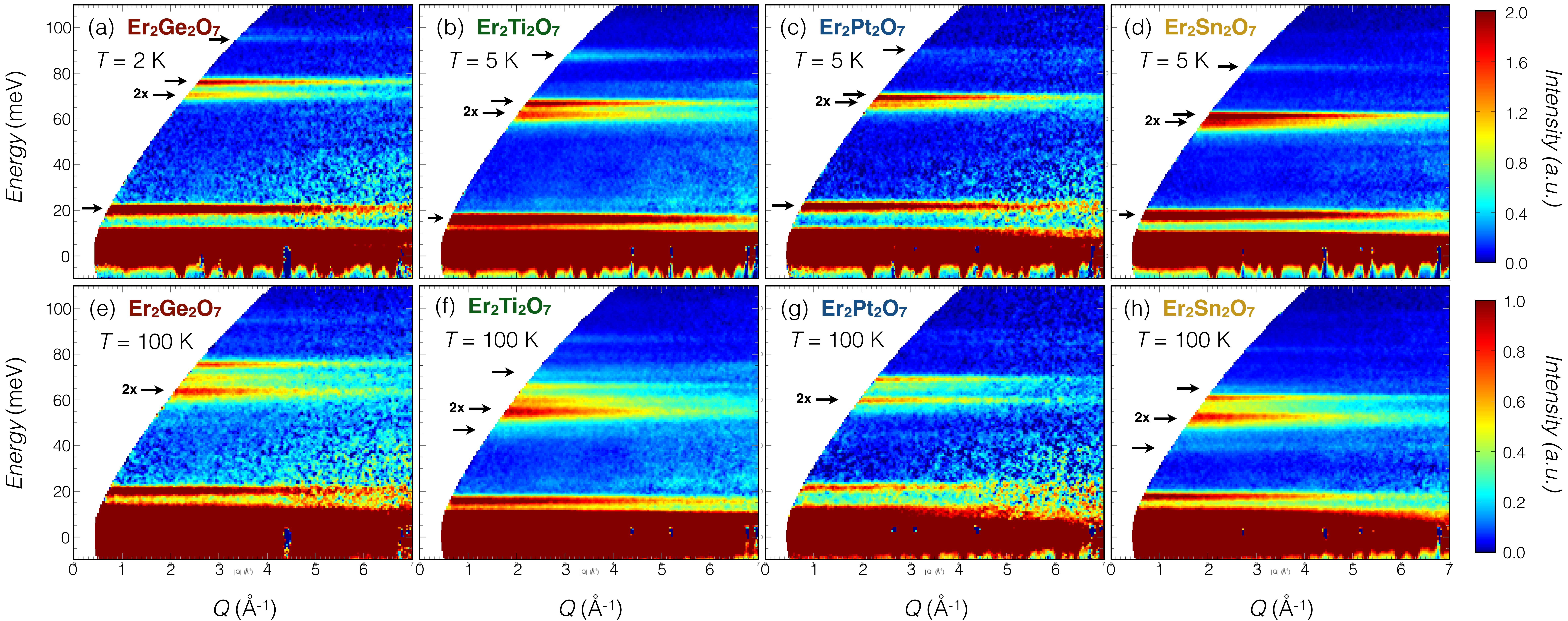}
\par
\caption{Inelastic neutron scattering measurements with an incident energy of $E_i = 150$~meV for the Er$_2B_2$O$_7$ pyrochlores, $B=$~(a,e) Ge, (b,f) Ti, (c,g) Pt, and (d,h) Sn. The top row shows the spectra at base temperature (2~K or 5~K) for each sample. In each case, four higher energy ground state crystal electric field excitations are identified at the positions indicated by the arrows, in addition to the transition at $\sim20$~meV, which was also seen in the $E_i=25$~meV data set. The bottom row show the spectra at 100~K, which is high enough to thermally populate transitions from the first and second excited state levels. The excited state transitions at 100~K are indicated by the arrows. An empty can measurement has been subtracted from all data sets.}
\label{150meV_Slices}
\end{figure*}

In order to characterize the structural properties of each of the four erbium pyrochlore magnets, we performed time-of-flight powder neutron diffraction. The data were analyzed by Rietveld refinement using the $Fd\bar{3}m$ pyrochlore space group, as shown in Figure~\ref{Refinement}. The fitted parameters resulting from these refinements are summarized in Table~\ref{RietveldTable}. The largest structural differences going across this series arise from the substitution of the non-magnetic $B$-site. Indeed the cubic lattice parameter, $a$, is observed to vary linearly with the ionic radius of the $B$-site cation. Thus, Er$_2$Ge$_2$O$_7$, which has the smallest $B$-site cation, has a lattice parameter of $a=9.87(5)$~\AA, while Er$_2$Sn$_2$O$_7$ with the largest $B$-site, has a lattice parameter of $a = 10.30$~\AA. Across the whole series, the lattice parameter varies by 4\%. It is interesting to note that the two middle members, Er$_2$Ti$_2$O$_7$ and Er$_2$Pt$_2$O$_7$, have the most similar $B$-site radii and correspondingly, the most similar lattice parameters, varying by only 0.7\%. Er$_2$Pt$_2$O$_7$, however, is a unique member of this family due to the fact that platinum, unlike the other non-magnetic $B$-site ions, does not have a closed electron shell configuration. Platinum in its 4+ oxidation state has an [Xe]$5d^6$ electron configuration, which is non-magnetic due to the filled $t_{2g}$ levels and sizable gap to the empty $e_g$ levels.\\ 

In the cubic pyrochlore structure, erbium sits at the 16$d$ Wyckoff position and is surrounded by a distorted cube of oxygen anions. This distortion acts along the local $\ev{111}$ direction, compressing the cube along the body diagonal. Thus, there are two equivalent Er-O1 bonds and six equivalent Er-O2 bonds. The degree of this distortion is characterized by the oxygen (O1) $x$ coordinate, which is the only adjustable parameter in the pyrochlore structure. An oxygen environment taking up a perfect cube would have $x$ = 0.375, and the distortion grows with the deviation from this value. The distortion from an ideal cubic environment can alternatively be parameterized by the ratio of the Er-O1 and Er-O2 bond lengths, as tabulated in Table~\ref{RietveldTable}. The most distorted erbium oxygen environment occurs in Er$_2$Ge$_2$O$_7$ and the least distorted environment occurs in Er$_2$Pt$_2$O$_7$. Aside from Er$_2$Pt$_2$O$_7$, this distortion scales linearly with lattice parameter. Lastly, it is worth mentioning that the erbium-oxygen bond distances are generally the shortest in Er$_2$Ge$_2$O$_7$ ($\langle\text{Er}-\text{O}\rangle = 2.21$~\AA) and largest in Er$_2$Sn$_2$O$_7$ ($\langle\text{Er}-\text{O}\rangle = 2.29$~\AA).\\

\subsection{Determination of the crystal field Hamiltonian with inelastic neutron scattering}

\begin{figure}[tbp]
\linespread{1}
\par
\includegraphics[width=3.2in]{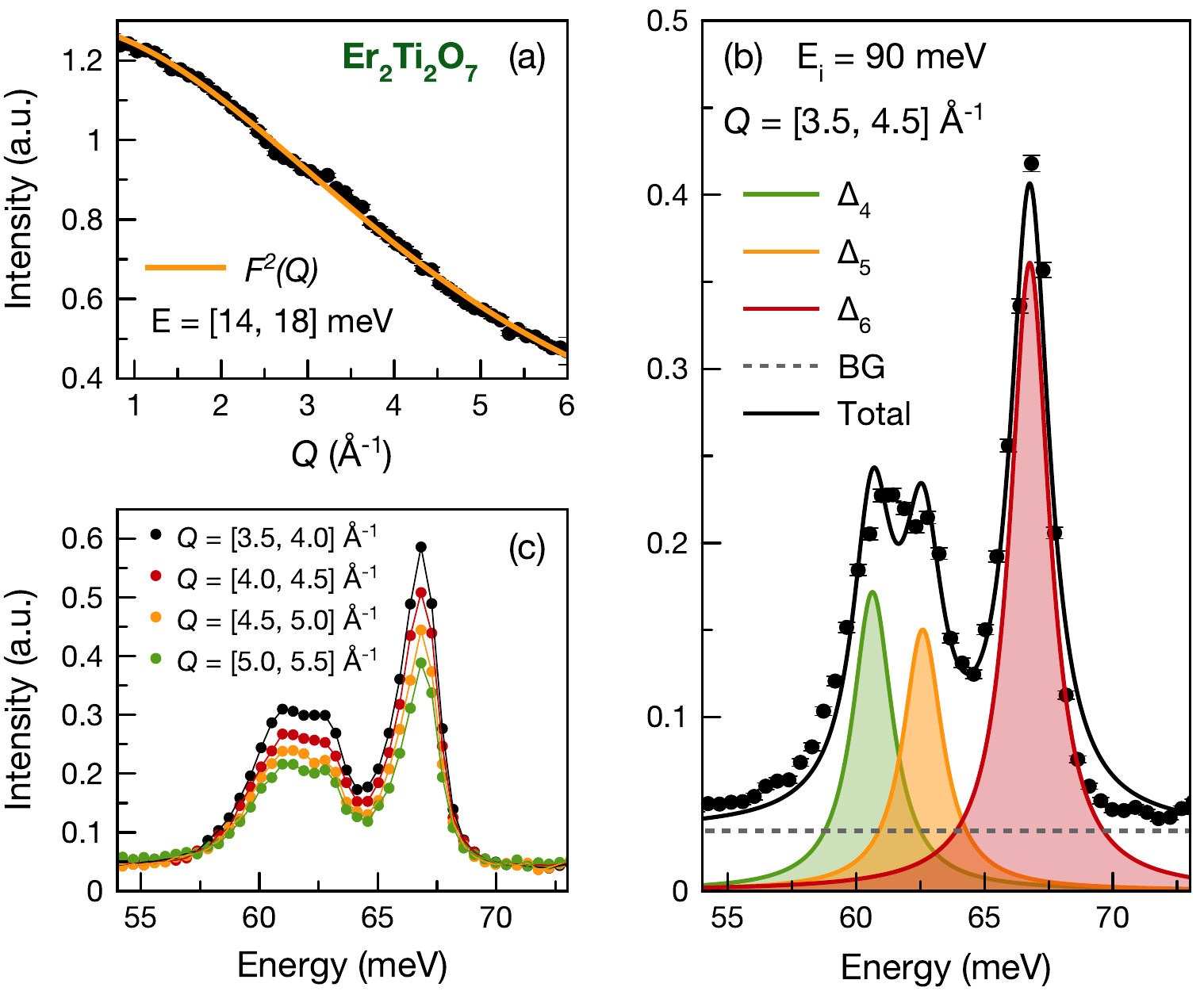}
\par
\caption{(a) The $|\vec{Q}|$ dependence of the 15.7~meV CEF excitation of Er$_2$Ti$_2$O$_7$ is well captured by the single ion magnetic form factor of Er$^{3+}$. The same $|\vec{Q}|$ dependence is observed for all the CEF excitations resolved in this work. (b) The neutron scattering intensity as a function of energy transfer for Er$_2$Ti$_2$O$_7$, revealing the presence of three CEF excitations over the energy range shown. An example of a fit to the CEF spectra is shown. The fit captures the integrated intensity for all three CEF excitations and was used to optimize the CEF Hamiltonian. (c) The $|\vec{Q}|$ dependence associated with the Er$^{3+}$ magnetic form factor can also be seen by performing energy cuts of the data with integrations over different $|\vec{Q}|$ ranges.}
\label{Splitting}
\end{figure}

The crystal electric field (CEF) spectra of the four erbium pyrochlore magnets were measured with inelastic neutron scattering. Data sets employing incident energies of 25~meV and 150~meV were collected at $T=2$ or 5~K for all samples, and are shown in the upper panels of Fig.~\ref{25meV_Slices} and Fig.~\ref{150meV_Slices}. As CEF excitations are single ion properties, they largely lack dispersion and decrease in intensity as a function of the momentum transfer ($|\vec{Q}|$) following the magnetic form factor of Er$^{3+}$. From these criteria, it is clear that the $E_i=25$~meV data sets at base temperature in Fig.~\ref{25meV_Slices}(a-d) each show three CEF excitations, as indicated by the black arrows. For example, Er$_2$Ti$_2$O$_7$ has CEF excitations at 6.3, 7.3 and 15.7~meV. An example of the $|\vec{Q}|$ dependence of the CEF scattering is shown in Fig.~\ref{Splitting}(a) for the 15.7~meV excitation in Er$_2$Ti$_2$O$_7$, where the solid line corresponds to the magnetic form factor of Er$^{3+}$. All other CEF transitions observed in this work follow a similar $|\vec{Q}|$ dependence. As will be discussed later, it is interesting to note that the lack of dispersion of the low energy CEF excitations is not complete. Weak dispersion in the lowest energy levels develops at low temperatures as a consequence of exchange coupling between the Er$^{3+}$ moments.\\

For each of the erbium pyrochlores studied in this work, the energy of the first CEF excitation is $\sim$6~meV above the ground state. As these measurements were performed at $T=2$ or 5~K, this first excited state cannot be significantly thermally populated and thus all observed CEF excitations necessarily originate from the CEF ground state. As previously discussed, the lowest energy manifold for Er$^{3+}$ with $J = 15/2$ is composed of 8 doublets, one of which is the ground state. Thus, to resolve the full CEF manifold requires the identification of a total of seven transitions. Turning our attention to the base temperature $E_i=150$~meV data sets shown in Fig.~\ref{150meV_Slices}(a-d), three additional CEF transitions are clearly observed for all samples, as indicated by the black arrows. For example, Er$_2$Ti$_2$O$_7$ has CEF transitions at 61.0, 66.3 and 87.2~meV. From our analysis thus far of the $E_i=25$~meV and $E_i=150$~meV data sets, we can account for three CEF levels below 20~meV and three additional CEF levels between 50~meV and 100~meV.\\

\begin{figure}[tbp]
\linespread{1}
\par
\includegraphics[width=3.2in]{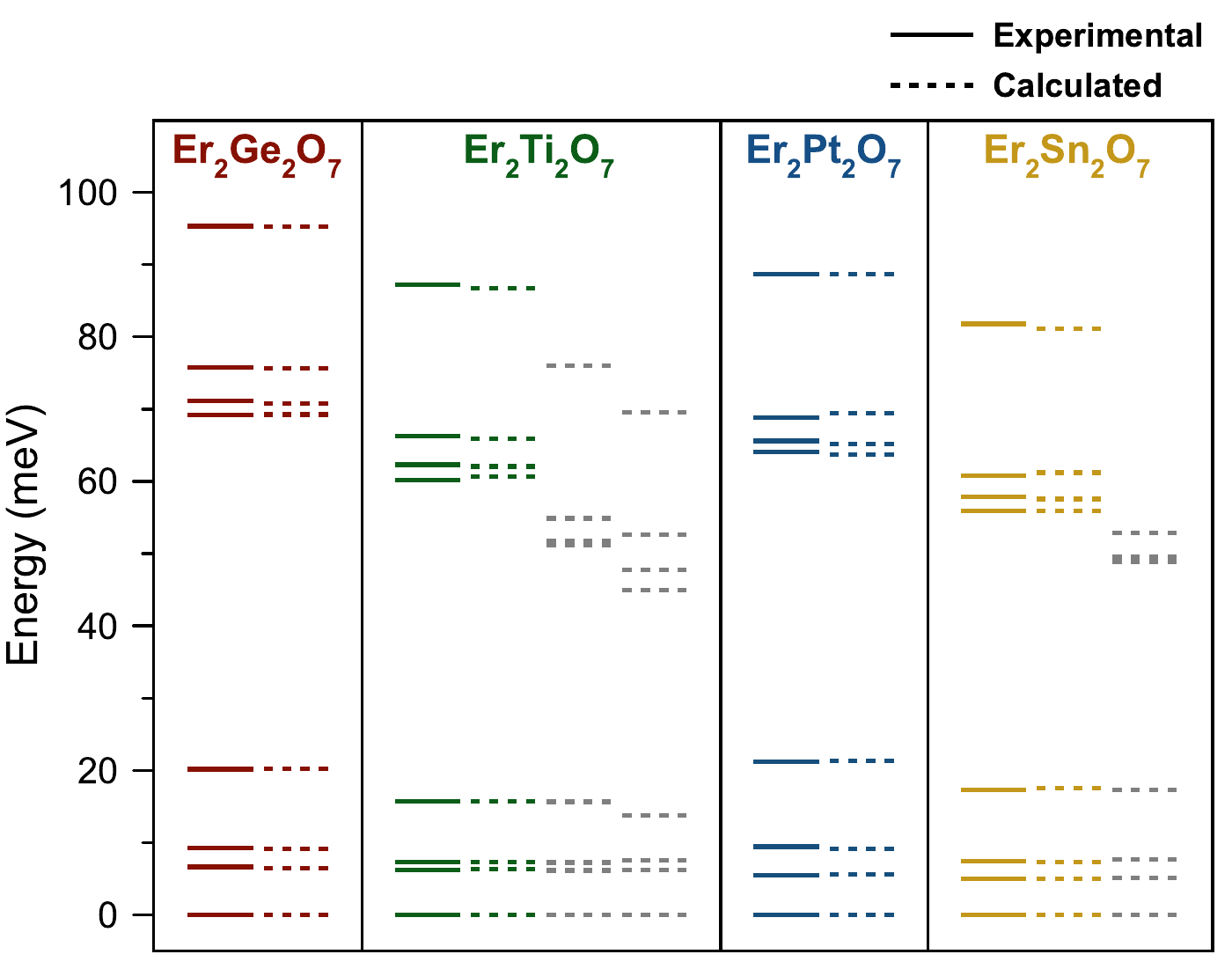}
\par
\caption{Comparison between the experimentally resolved CEF energy schemes of the four erbium pyrochlore magnets and those calculated using the CEF Hamiltonians given in Table~\ref{CEF_Solutions_Table}. The calculated CEF spectra for Er$_2$Ti$_2$O$_7$~\cite{bertin2012crystal,GuittenyErSnO} and Er$_2$Sn$_2$O$_7$~\cite{GuittenyErSnO} derived from previous studies are also compared to our experimental data and are given by the gray dashed lines.}
\label{EnergyScheme}
\end{figure}

As seven transitions are required to fully determine the energy scheme of the CEF manifold, there is a final CEF excitation that is not immediately resolved in the $E_i=25$~meV and $E_i=150$~meV data sets. However, inelastic neutron scattering spectra collected with an incident energy of 90~meV allow us to resolve the missing CEF level. For Er$_2$Ti$_2$O$_7$, the level centered near 61.0~meV, as shown in Fig.~\ref{Splitting}(b), is in fact, two closely spaced CEF transitions that were unresolvable using higher incident energy with corresponding lower energy resolution. Fig.~\ref{Splitting}(c) shows the decreasing intensity of these two peaks as a function of momentum transfer, confirming that both features are magnetic in origin and thus, CEF excitations. The corresponding CEF level in each of the Er pyrochlore magnets is split by a similar amount. With this final CEF transition now accounted for, the entire CEF manifold is known and is presented in Fig.~\ref{EnergyScheme} for the four samples probed in this work.\\

\begin{figure*}[tbp]
\linespread{1}
\par
\includegraphics[width=6.5in]{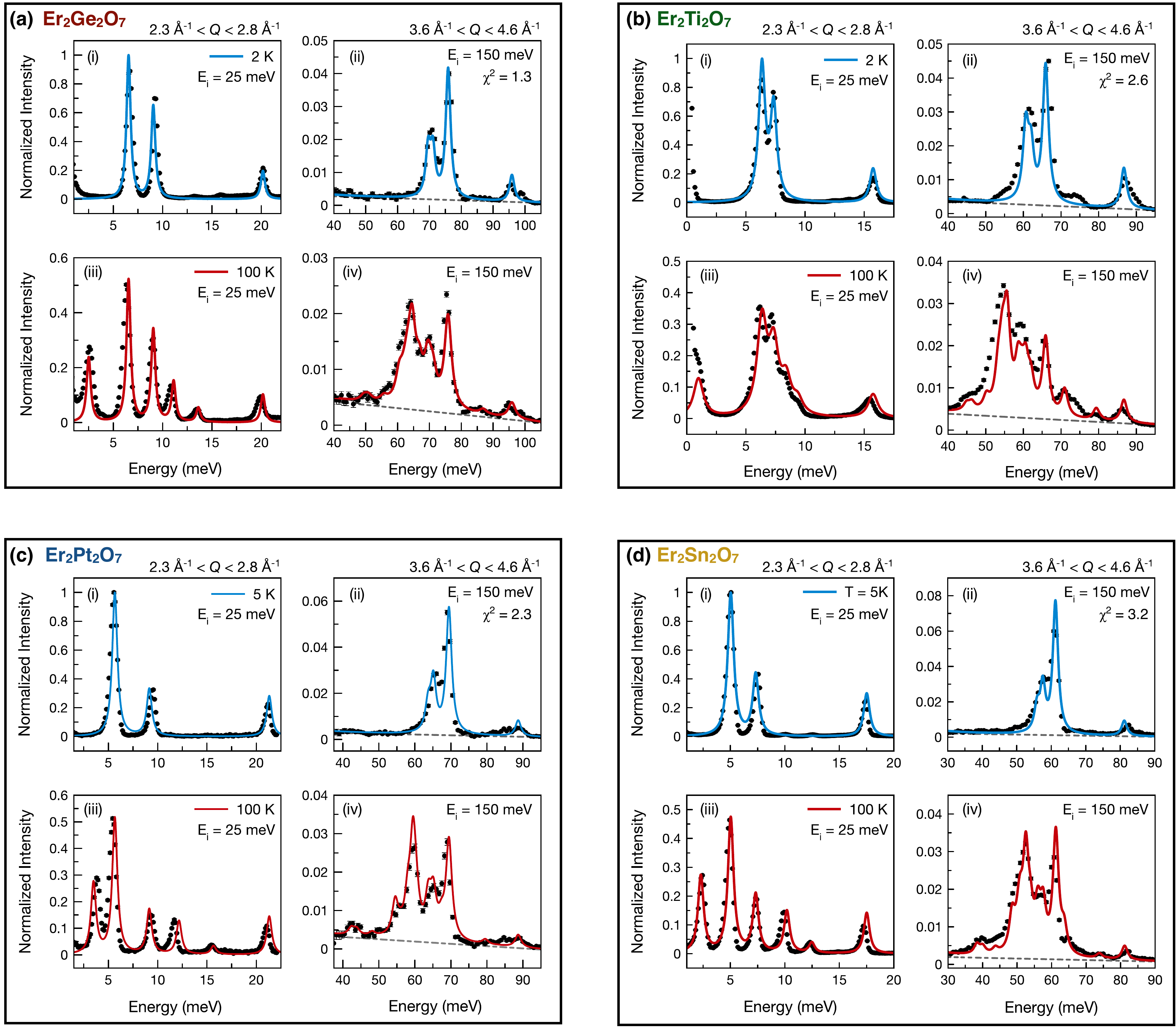}
\par
\caption{Inelastic neutron spectra for (a) Er$_2$Ge$_2$O$_7$, (b) Er$_2$Ti$_2$O$_7$, (c) Er$_2$Pt$_2$O$_7$ and (d) Er$_2$Sn$_2$O$_7$ showing the scattering intensity as a function of energy transfer at base temperature ((i) and (ii) and 100K ((iii) and (iv)). These data sets are obtained from the contour maps of Fig.~\ref{25meV_Slices} and Fig.~\ref{150meV_Slices} with an integration in momentum transfer, $|\vec{Q}|$ from 2.3 to 2.8~$\angstrom^{-1}$ for (i) and (iii) and from 3.6 to 4.6~$\angstrom^{-1}$for (ii) and (iv). The calculated CEF spectra using the CEF Hamiltonian parameters shown in Table~\ref{CEF_Solutions_Table} are plotted by the blue ($T=2$ and 5~K) and red ($T=100$~K) lines.}
\label{CEF_Fits}
\end{figure*}

Our experimentally deduced CEF scheme can be compared to the CEF Hamiltonians previously derived for Er$_2$Ti$_2$O$_7$~\cite{bertin2012crystal,GuittenyErSnO} and Er$_2$Sn$_2$O$_7$~\cite{GuittenyErSnO}. Using the parametrization from these previous works we have computed the CEF energy schemes, which are also presented in Fig.~\ref{EnergyScheme}. By comparing with our new results, it is clear that the previous CEF Hamiltonians do not reproduce well the high energy CEF levels. Thus, a new refinement of the CEF Hamiltonians was performed for Er$_2$Ti$_2$O$_7$ and Er$_2$Sn$_2$O$_7$ as well as for Er$_2$Ge$_2$O$_7$ and Er$_2$Pt$_2$O$_7$, for which no previous CEF Hamiltonian has been derived. With the new results presented in this work, it is now possible to rigorously constrain the CEF Hamiltonian because the full spin-orbit ground state manifold has been experimentally determined. This is in contrast to previous works that only considered the low energy excited states. To execute the fitting procedure, integrations of the scattered intensity as a function of energy transfer between $Q$ = 2.3 to 2.8~$\angstrom^{-1}$ for the $E_i=25$~meV data set and $Q$ = 3.6 to 4.6~$\angstrom^{-1}$ for the $E_i=150$~meV data set have been extracted from the base temperature contour maps of Fig.~\ref{25meV_Slices} and Fig.~\ref{150meV_Slices}. The energy of each CEF level was determined by fitting the transition with a Lorentzian function for which the area represents the relative scattered intensity of the particular CEF transition. The relative intensities of the  CEF transitions measured with the 25~meV, 90~meV and 150~meV were normalized to each other using the third-excited CEF state which appeared in all data sets (for example, the CEF at 20.2~meV for Er$_2$Ge$_2$O$_7$). In this manner, the CEF Hamiltonian was constrained using the energy of each transition and their relative scattered intensity, giving a total of 13 constraints, for each of the four erbium pyrochlore magnets studied.\\

\begin{table}[]
\centering
\caption{The refined CEF parameters $A_n^m$ obtained for all four erbium pyrochlore magnets studied in this work.}
\label{CEF_Solutions_Table}
\begin{tabular}{lp{1cm}<{\centering}p{1cm}<{\centering}p{1cm}<{\centering}p{0.8cm}<{\centering}p{1cm}<{\centering}p{1cm}<{\centering}}
\hline
\hline
\multicolumn{1}{r}{\textit{(all in meV)}} & A$^0_2$ & A$^0_4$ & A$^3_4$ & A$^0_6$ & A$^3_6$  & A$^6_6$ \\ \hline
Er$_2$Ge$_2$O$_7$                         & 39.3   & 36.2  & 275   & 1.23    & $-19.10$  & 26.6    \\
Er$_2$Ti$_2$O$_7$                         & 37.5    & 33.5    & 282   & 1.25    & $-17.15$ & 21.6    \\
Er$_2$Pt$_2$O$_7$                         & 49.0    & 31.2    &  262 & 1.10     &  $-19.15$  & 24.6        \\
Er$_2$Sn$_2$O$_7$                         & 50.8  & 28.4  & 266   &  1.10    & $-17.15$ & 20.9   \\ \hline
\hline
\end{tabular}
\end{table}

Using the extracted experimental constraints, the CEF Hamiltonian was first diagonalized using the CEF parameters of Bertin \emph{et al.} derived for Er$_2$Ti$_2$O$_7$~\cite{bertin2012crystal}. A broad scan of the CEF parameters around this particular solution was performed to minimize the $\chi^2$ value between the calculated and experimental CEF schemes, which considers both the energies of the transitions as well as their relative intensities. Very good agreement was obtained for all samples using the appropriate CEF parameters presented in Table~\ref{CEF_Solutions_Table}. The results of this optimization procedure are presented in Table~\ref{Exp_value} of the Appendix, where the computed energies and intensities are compared to the experimental values. The resulting $\chi^2$ of our calculation is equal to 1.3, 2.6, 2.3 and 3.2 for Er$_2$Ge$_2$O$_7$, Er$_2$Ti$_2$O$_7$, Er$_2$Pt$_2$O$_7$ and Er$_2$Sn$_2$O$_7$ respectively. To better appreciate the results of the fitting procedure, the entire neutron scattering spectra were computed between 0 and 120~meV using the CEF parameters of Table~\ref{CEF_Solutions_Table} with the addition of a sloping background to account for the phonon contribution. The resulting theoretical calculation is compared to the data in Fig.~\ref{CEF_Fits} for the four erbium pyrochlore samples. Excellent agreement between the calculated and measured scattering spectra is observed with no significant differences, validating the goodness of the fits.\\

\begin{table*}[]
\centering
\caption{The composition of the CEF ground state doublet for the four erbium pyrochlore magnets studied in this work. The calculated components of the $g$-tensor perpendicular ($g_{\perp}$) and parallel ($g_z$) to the local $\ev{111}$ axis are also shown along with their ratio $\sfrac{g_{\perp}}{g_z}$ and the calculated moment ($\mu_{cef}$). The experimentally determined magnetic ground state (G.S) and its associated magnetic moment ($\mu_{ord}$) obtained via powder neutron diffraction are also given for each erbium pyrochlore.}
\label{GS_Eigenfunctions_Table}
\begin{tabular}{lp{1.3cm}<{\centering}p{1.3cm}<{\centering}p{1.3cm}<{\centering}p{1.3cm}<{\centering}p{1.3cm}<{\centering}p{1.1cm}<{\centering}p{1.1cm}<{\centering}p{1.1cm}<{\centering}p{1.5cm}<{\centering}p{1.4cm}<{\centering}p{1.5cm}<{\centering}}
\hline
\multicolumn{1}{r}{\textit{}} & $\ket{\pm13/2}$ & $\ket{\pm7/2}$ & $\ket{\pm1/2}$ & $\ket{\mp5/2}$ & $\ket{\mp11/2}$ & $g_{\perp}$ & $g_z$ & $\sfrac{g_{\perp}}{g_z}$  & $\mu_{cef}$ ($\mu_B$) & $\mu_{ord}$ ($\mu_B$) & G.S \\ \hline
Er$_2$Ge$_2$O$_7$  & 0.395 & 0.338  & $-0.521$  & $-0.176$  & 0.653  & 7.0 & 2.1 & 3.3 &  3.6 & 3.23(6) & $\Gamma_5$~\cite{Dun2015}  \\
Er$_2$Ti$_2$O$_7$  & 0.389 & 0.257  & $-0.491$  & $-0.082$  & 0.732  & 6.3 & 3.9 & 1.6 &  3.7 & 3.25(9) & $\Gamma_5$~\cite{Poole2007} \\
Er$_2$Pt$_2$O$_7$  & 0.381 & 0.439  & $-0.563$  & $-0.280$  & 0.516  & 7.7 & 0.3 & 28  &  3.9 & 3.4(2) & $\Gamma_7$~\cite{Hallas2017EPO}  \\
Er$_2$Sn$_2$O$_7$  & 0.541 & 0.244  & $-0.578$  & $-0.413$  & 0.379  & 7.6 & 0.14 & 54  &  3.8 & 3.1 & $\Gamma_7$~\cite{Petit2017}  \\ \hline
\end{tabular}
\end{table*}

To further scrutinize the quality of our CEF Hamiltonian parameters, inelastic neutron scattering spectra were collected at a temperature of 100~K, using both $E_i = 25$~meV and $E_i=150$~meV, which is shown in the bottom panels of Fig.~\ref{25meV_Slices} and Fig.~\ref{150meV_Slices}. The increase of temperature from 5~K to 100~K produces new CEF excitations that originate from the increasing thermal population of the first and second excited CEF states. The new (relative to low temperature) low energy CEF excitations are indicated by the black arrows in the bottom panels of Fig.~\ref{25meV_Slices}. For example, new transitions are observed at 1.1, 8.4 and 9.5~meV in Er$_2$Ti$_2$O$_7$ at 100~K. These originate from transitions between the 1st and 2nd excited CEF states, the 1st and 3rd excited CEF states, and the 2nd and 3rd excited CEF states. For the 100~K high energy spectra shown in Fig.~\ref{150meV_Slices}, multiple new CEF excitations are observed between 40 and 100~meV. These also correspond to transitions from the first and second excited CEF states to higher energy CEF states. The prominent new CEF transitions are indicated by black arrows in the bottom panels of Fig.~\ref{150meV_Slices}.\\

Using Equation~2, we can compute the predicted inelastic scattering spectra at 100~K using the CEF parameters derived from our base temperature fits, given in Table~\ref{CEF_Solutions_Table}. The resulting calculations can be compared with the experimental data by performing cuts in energy integrated over the same range in $|\vec{Q}|$ as previously performed on the base temperature data sets. A sloping background has been added to the calculation to account for the increased phonon contribution. The comparison between the theoretical curves and the experimental data at 100~K are shown in Fig.~\ref{CEF_Fits} for all four samples. The computed spectra is entirely consistent with the experimental data, demonstrating that our CEF Hamiltonian fits are well constrained and further validating the quality of the fit. Note that by comparing the 2~K and 100~K experimental spectra, slight negative energy shifts of the CEF transitions originating from the CEF ground state are observed upon raising the temperature. Such softening of the CEF excitations has also been observed in other rare-earth pyrochlore systems~\cite{gaudet2015neutron}, and this is due to lattice expansion that occurs on raising the temperature, which in turn reduces the strength of the electric field at the Er$^{3+}$ position.\\

\section{Discussion}

With their CEF Hamiltonians in hand, it is interesting to analyze the evolution of the single-ion properties going across the erbium pyrochlore series. The composition of the CEF ground state doublet and the associated $g$-tensor components for each of the Er$_2B_2$O$_7$ materials are shown in Table~\ref{GS_Eigenfunctions_Table}. Across the family, the compositions of the ground state doublets are qualitatively similar and consist of a mixture of $J_z=\ket{\pm\frac{13}{2}},\ket{\pm\frac{7}{2}},\ket{\pm\frac{1}{2}},\ket{\mp\frac{5}{2}},\ket{\mp\frac{11}{2}}$. In each case, the ground state doublet has local XY-like anisotropy because the $g$-tensor perpendicular to the local $\ev{111}$ ($g_{\perp}$) is larger than the parallel ($g_z$) component. However, the strength of the local XY anisotropy is an order of magnitude smaller for Er$_2$Ge$_2$O$_7$ and Er$_2$Ti$_2$O$_7$ than it is for Er$_2$Pt$_2$O$_7$ and Er$_2$Sn$_2$O$_7$. This result suggests that the change in the $g$-tensor anisotropy going across the family could underly the stabilization of different magnetically ordered states. Indeed, both Er$_2$Ge$_2$O$_7$ and Er$_2$Ti$_2$O$_7$ order into the $k=0$ $\Gamma_5$ manifold~\cite{Poole2007,Dun2015} and possess relatively weak XY anisotropy. In contrast, both Er$_2$Pt$_2$O$_7$ and Er$_2$Sn$_2$O$_7$ order into the $k=0$ $\Gamma_7$ manifold~\cite{Hallas2017EPO,Petit2017} and possess strong XY anisotropy. 

To examine such a hypothesis, we can refer to the anisotropic exchange phase diagram of Ref.~\cite{wong2013ground}, where the four exchange couplings are labeled $J_{zz}$, $J_{\pm}$, $J_{\pm\pm}$, and $J_{z\pm}$. The classical phase boundary between the $\Gamma_5$ and $\Gamma_7$ states occurs at a critical ratio of $\frac{J_{\pm\pm}}{J_{\pm}}=2$. Above this value, the predicted ground state is $\Gamma_7$ and below this value, the predicted ground state is $\Gamma_5$~\cite{wong2013ground}. Interestingly, both $J_{\pm\pm}$ and $J_{\pm}$ are proportional to $g_{\perp}^2$~\cite{GuittenyErSnO}, and thus, the ratio of $\frac{J_{\pm\pm}}{J_{\pm}}$ should not depend on the $g$-tensor anisotropy per se. This would then suggest that the transition in magnetic ground state across the erbium pyrochlore family is driven by changes in the details of the orbital overlap and not the single-ion properties. Related to this point, it is interesting to note that the oxygen environment derived from our powder neutron analysis reveals a more cubic environment for Er$_2$Pt$_2$O$_7$ and Er$_2$Sn$_2$O$_7$ as compared to Er$_2$Ge$_2$O$_7$ and Er$_2$Ti$_2$O$_7$. This may indicate a more isotropic exchange interaction for the two former materials, consistent with recent estimates of the exchange parameters for Er$_2$Sn$_2$O$_7$~\cite{GuittenyErSnO,Petit2017}. Nonetheless, the two members of the erbium pyrochlore family that order into the Palmer-Chalker state also show extreme XY anisotropy, very small $g_z$, and a more cubic local environment around the Er$^{3+}$ site as compared with the two members that order into $\Gamma_5$ states. \\

\begin{figure}[tbp]
\linespread{1}
\par
\includegraphics[width=3.2in]{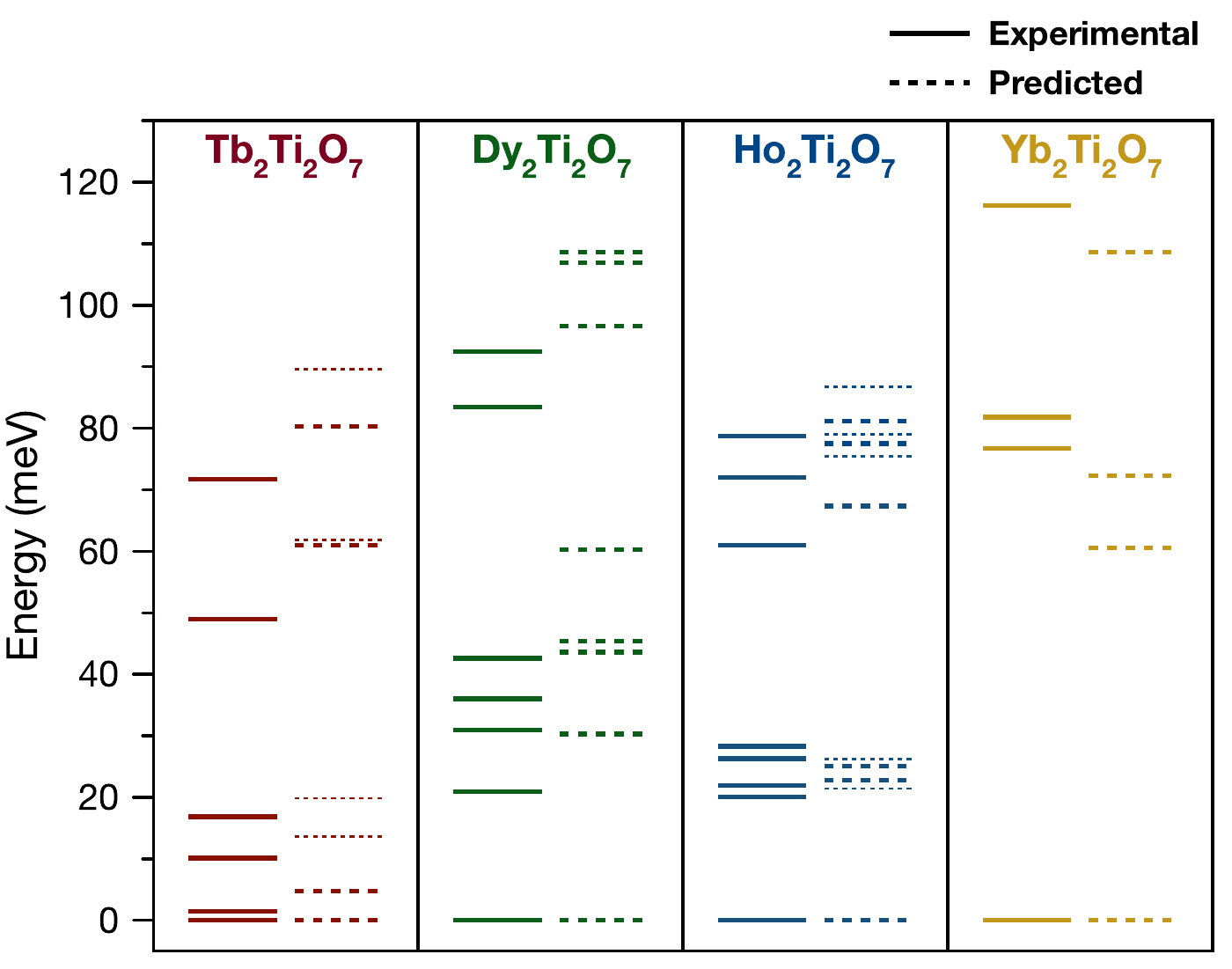}
\par
\caption{Calculated CEF schemes for the rare earth titanates, $A_2$Ti$_2$O$_7$ ($A=$~Tb, Dy, Ho, and Yb) obtained using scaling arguments from our fitted parameters for Er$_2$Ti$_2$O$_7$. The calculated levels are indicated by the dashed lines where, for Tb$_2$Ti$_2$O$_7$ and Ho$_2$Ti$_2$O$_7$, the singlet levels are denoted by a lighter dashed line. The experimentally determined CEF energies, reproduced from Refs.~\cite{rosenkranz2000crystal,gaudet2015neutron,ruminy2016crystal}, are given by the solid lines.}
\label{CEF_otherpyro}
\end{figure}

This new study of the erbium pyrochlores has resolved all possible CEF excitations within the lowest energy $J$ multiplet, an analysis that has not been achieved for any other titanate pyrochlore besides Yb$_2$Ti$_2$O$_7$~\cite{gaudet2015neutron}, which only has three excited levels. Furthermore, as Er$^{3+}$ is a Kramers ion, the assignment of each CEF level as a doublet is unambiguous, in contrast to non-Kramers Tb$_2$Ti$_2$O$_7$ and Ho$_2$Ti$_2$O$_7$. It is therefore informative to use a scaling argument to approximate the CEF schemes for other pyrochlore magnets based on our new comprehensive results. As the titanates are the best studied family of insulating rare earth pyrochlores, we have focused on approximating the CEF schemes for $A_2$Ti$_2$O$_7$ ($A=$~Tb, Dy, Ho, and Yb) starting from the parameters obtained for Er$_2$Ti$_2$O$_7$. The scaling argument that connects the CEF parameters, $A_n^m(R)$, between different rare earth-based pyrochlores is given by the following~\cite{Hutchings}:
\begin{eqnarray}
A_n^m(R')=\frac{a^{n+1}(R)}{a^{n+1}(R')}A_n^m(R),
\end{eqnarray} 
where the cubic lattice parameters, $a(R)$, have been taking from Ref.~\cite{li2013single}. The resulting CEF Hamiltonian approximation is given in Fig.~\ref{CEF_otherpyro} where it is compared with the measured schemes from Refs.~\cite{rosenkranz2000crystal,gaudet2015neutron,ruminy2016crystal}. The CEF schemes obtained from this procedure have good qualitative agreement with the experimental data, but are quantitatively inaccurate, which should not be surprising for several reasons. First, this analysis assumes that the oxygen environment surrounding each rare earth ion has the same O1 $x$ parameter and that the oxygen cage simply scales with the lattice parameter. However, it is known that the distortion of the oxygen cage depends on the $A$-site ion~\cite{lau2006stuffed}, which directly impacts the CEF energy scheme. Second, the shielding factor, which influences the overall scale of the CEF splitting, also depends on the $A$-site ion. This second point explains most of the discrepancy between our calculation and the experimental data. The correct form of the CEF scheme is observed for all the calculated spectra, but the overall energy range is either too high (for $A=$~Tb, Dy, and Ho) or too low (for $A=$~Yb). In the cases where the scaling is too large, the cation in question lies to the left of erbium on the periodic table and the opposite is true when the scaling is too small. The best agreement is obtained for Ho$_2$Ti$_2$O$_7$, where Er and Ho are next-door neighbors in the periodic table. These observations suggest that the shielding factor decreases on going from smaller (heavier) to bigger (lighter) rare earth ions.\\

\begin{figure}[tbp]
\linespread{1}
\par
\includegraphics[width=3.2in]{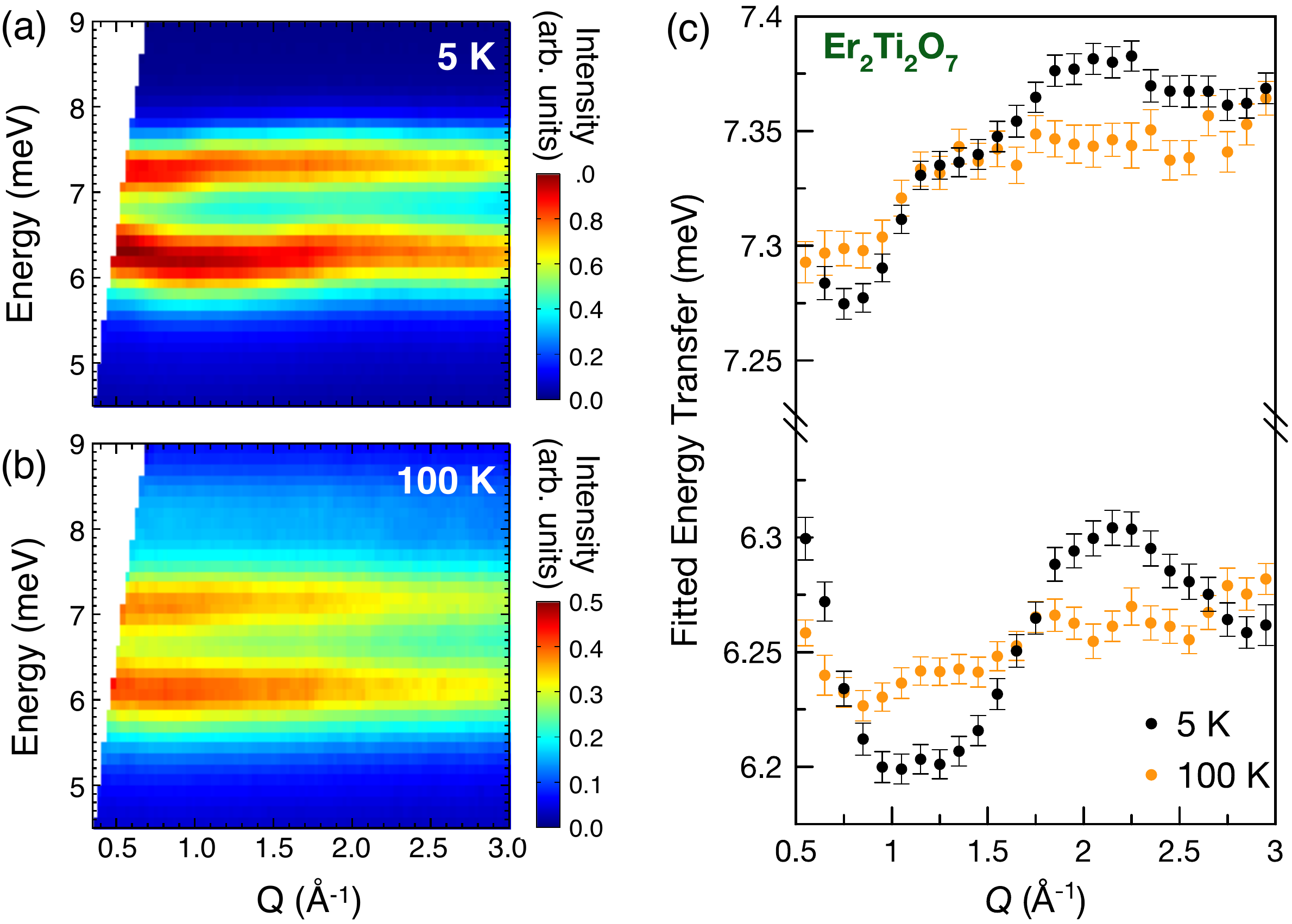}
\par
\caption{An enhanced view of the lowest energy crystal electric field levels in Er$_2$Ti$_2$O$_7$ at (a) 5~K and (b) 100~K, revealing weak dispersion. (c) The maximum fitted intensity of these two CEF excitations as a function of $|\vec{Q}|$ at both 5~K and 100~K.}
\label{CEF_dispersion}
\end{figure}

Even if a quantitative discrepancy exists between the real and calculated CEF schemes, the scaling procedure can still be useful in terms of providing a starting point for analyzing the CEF scheme of any pyrochlore oxide. For example, it has been observed that the low energy CEF scheme for Tb$_2$Ti$_2$O$_7$ consists of four distinct magnetic excitations at approximatively 1, 10, 14 and 16~meV~\cite{zhang2014neutron,princep2015crystal,ruminy2016crystal}. However, no CEF Hamiltonian has been able to capture such a CEF energy scheme while remaining consistent with the high energy levels. Rather, one excited CEF doublet at 1~meV and two excited CEF singlets at 10 and 16~meV have been refined~\cite{princep2015crystal,ruminy2016crystal}. To resolve this issue, it has been proposed that spin-phonon coupling is the origin of the extra inelastic feature observed at 14~meV~\cite{ruminy2016crystal}. Our CEF approximation for Tb$_2$Ti$_2$O$_7$, based on the scaling arguments above, also reveals only one excited CEF doublet at 4~meV and two excited CEF singlets at 12 and 20~meV. Thus, our scaling calculation further validates the spin-phonon origin of the enigmatic fourth excitation observed in Tb$_2$Ti$_2$O$_7$ below 20~meV.\\    

Finally, as briefly alluded to earlier, our neutron scattering results indicate that the low energy CEF excitations in the erbium pyrochlores do not have a perfectly flat dispersion. To illustrate this effect, an enhanced view on the lowest energy CEF excitations of Er$_2$Ti$_2$O$_7$ measured with $E_i=$25~meV is shown in Fig.~\ref{CEF_dispersion}(a). This reveals weak dispersion for the CEF excitation at 6.25~meV and possibly also for the CEF excitation at 7.3~meV. As the temperature is increased to $T=100$~K, this dispersion becomes much less pronounced as shown in Fig.~\ref{CEF_dispersion}(b). To quantify the dispersion, we performed integrations along different $|\vec{Q}|$ values with a width of 0.1~$\angstrom^{-1}$. For every integration, we determined the positions of the low energy CEF excitations by fitting their line-shapes to a Gaussian function. The fitted centers of the Gaussian are plotted as a function of $|\vec{Q}|$ in Fig.~\ref{CEF_dispersion}(c) for both 5~K and 100~K. At 5~K, the 6.25~meV CEF excitation has a minimum in energy at $|\vec{Q}|$ $\sim$ 1.1~$\angstrom^{-1}$, which corresponds to the range of $|\vec{Q}|$ where the (111) magnetic Bragg peak develops at $T_N=1.2$~K in Er$_2$Ti$_2$O$_7$. The bandwidth of the dispersion of the 6.25 meV CEF excitation is 0.1~meV at 5 K, and perhaps a little less for the CEF excitation near 7.3 meV. At 100~K, the dispersion of the 6.25~meV CEF excitation is markedly decreased and it no longer exhibits a clear minimum at $|\vec{Q}|$ $\sim$ 1.1~$\angstrom^{-1}$. The temperature change in the dispersion of the CEF at 7.3~meV is less dramatic than the CEF at 6.25~meV, but the same effect is observed - less dispersion at higher temperature.\\

Weak dispersion is also observed for the first and second excited CEF levels of the three other erbium pyrochlores probed in this work. The shape of the dispersion, as well as its suppression at high temperature, are similar throughout the family. The origin of this weak dispersion is likely exchange interactions that lead to enhanced intersite correlations at low temperatures. Similar effects are observed in terbium pyrochlores, for which the lowest lying CEF excitations, located near 1~meV, display intense dispersion relative to their mean energy~\cite{gardner1999cooperative,gingras2000thermodynamic}. Virtual CEF transitions have been argued to contribute significantly to the low temperature magnetism of the terbium pyrochlore~\cite{molavian2007dynamically}. The magnitude of this effects scales with the inverse square of the energy gap between the ground state and the excited CEF state~\cite{rau2016order}. Despite the fact that the energy gap is a factor of approximately five larger in the erbium pyrochlores compared to the terbium pyrochlores, arguments have also been made that such virtual CEF transitions play a role in their ground state selection~\cite{mcclarty2009energetic,petit2014order,rau2016order}.\\ 

\section{Conclusion}

To conclude, we have carried out inelastic neutron scattering measurements on four members of the erbium pyrochlore family: Er$_2B_2$O$_7$ ($B=$~Ge, Ti, Pt, and Sn). These measurements allowed us to identify the entire set of CEF excitations belonging to the lowest energy $J$ multiplet for all four compounds. Weak dispersion of the lowest energy CEF excitation, with a softening near $|\vec{Q}|$ $\sim$ 1.1~$\angstrom^{-1}$, is observed at low temperatures. On the basis of these measurements, a CEF Hamiltonian was refined. These results give local XY-like anisotropy to the anisotropic $g$-tensors for all four erbium pyrochlore magnets. However, large differences in the degree of XY anisotropy were observed, with Er$_2$Ge$_2$O$_7$ and Er$_2$Ti$_2$O$_7$ displaying relatively more isotropic $g$-tensors as compared to Er$_2$Pt$_2$O$_7$ and Er$_2$Sn$_2$O$_7$, which display extreme XY anisotropy with very low values of $g_z$. This variation in the relative XY anisotropy correlates strongly with magnetic ground state selection, as Er$_2$Ti$_2$O$_7$ and Er$_2$Ge$_2$O$_7$ order into the $\Gamma_5$ antiferromagnetic structures, while Er$_2$Pt$_2$O$_7$ and Er$_2$Sn$_2$O$_7$ order into the $\Gamma_7$ Palmer-Chalker structures. Finally, our study represents the most detailed analysis of the single-ion properties of these erbium pyrochlores to date, a key ingredient for a detailed characterization of their spin Hamiltonians, and crucial for a clear understanding of the exotic magnetism observed in the XY pyrochlore magnets.\\  

\begin{acknowledgments}
We acknowledge useful conversations with K.A. Ross and J. G. Rau. We thank M. Tachibana, H. Sakurai, and H. Dabkowska for assistance in preparing the samples measured in this work. We thank Dr. Ashfia Huq and Dr. Melanie Kirkham of ORNL for collecting the neutron diffraction data. J.G. and A.M.H. acknowledges support from the Canada Graduate Scholarship Program. This research used resources at the Spallation Neutron Source, a DOE Office of Science User Facility operated by the Oak Ridge National Laboratory (ORNL). This work was supported by the Natural Sciences and Engineering Research Council of Canada.
\end{acknowledgments}

\bibliography{Er_CEF_Bib}

\begin{thebibliography}{49}%
\makeatletter
\providecommand \@ifxundefined [1]{%
 \@ifx{#1\undefined}
}%
\providecommand \@ifnum [1]{%
 \ifnum #1\expandafter \@firstoftwo
 \else \expandafter \@secondoftwo
 \fi
}%
\providecommand \@ifx [1]{%
 \ifx #1\expandafter \@firstoftwo
 \else \expandafter \@secondoftwo
 \fi
}%
\providecommand \natexlab [1]{#1}%
\providecommand \enquote  [1]{``#1''}%
\providecommand \bibnamefont  [1]{#1}%
\providecommand \bibfnamefont [1]{#1}%
\providecommand \citenamefont [1]{#1}%
\providecommand \href@noop [0]{\@secondoftwo}%
\providecommand \href [0]{\begingroup \@sanitize@url \@href}%
\providecommand \@href[1]{\@@startlink{#1}\@@href}%
\providecommand \@@href[1]{\endgroup#1\@@endlink}%
\providecommand \@sanitize@url [0]{\catcode `\\12\catcode `\$12\catcode
  `\&12\catcode `\#12\catcode `\^12\catcode `\_12\catcode `\%12\relax}%
\providecommand \@@startlink[1]{}%
\providecommand \@@endlink[0]{}%
\providecommand \url  [0]{\begingroup\@sanitize@url \@url }%
\providecommand \@url [1]{\endgroup\@href {#1}{\urlprefix }}%
\providecommand \urlprefix  [0]{URL }%
\providecommand \Eprint [0]{\href }%
\providecommand \doibase [0]{http://dx.doi.org/}%
\providecommand \selectlanguage [0]{\@gobble}%
\providecommand \bibinfo  [0]{\@secondoftwo}%
\providecommand \bibfield  [0]{\@secondoftwo}%
\providecommand \translation [1]{[#1]}%
\providecommand \BibitemOpen [0]{}%
\providecommand \bibitemStop [0]{}%
\providecommand \bibitemNoStop [0]{.\EOS\space}%
\providecommand \EOS [0]{\spacefactor3000\relax}%
\providecommand \BibitemShut  [1]{\csname bibitem#1\endcsname}%
\let\auto@bib@innerbib\@empty
\bibitem [{\citenamefont {Gardner}\ \emph {et~al.}(2010)\citenamefont
  {Gardner}, \citenamefont {Gingras},\ and\ \citenamefont {Greedan}}]{Greedan}%
  \BibitemOpen
  \bibfield  {author} {\bibinfo {author} {\bibfnamefont {J.~S.}\ \bibnamefont
  {Gardner}}, \bibinfo {author} {\bibfnamefont {M.~J.~P.}\ \bibnamefont
  {Gingras}}, \ and\ \bibinfo {author} {\bibfnamefont {J.~E.}\ \bibnamefont
  {Greedan}},\ }\href {\doibase 10.1103/RevModPhys.82.53} {\bibfield  {journal}
  {\bibinfo  {journal} {Rev. Mod. Phys.}\ }\textbf {\bibinfo {volume} {82}},\
  \bibinfo {pages} {53} (\bibinfo {year} {2010})}\BibitemShut {NoStop}%
\bibitem [{\citenamefont {Harris}\ \emph {et~al.}(1997)\citenamefont {Harris},
  \citenamefont {Bramwell}, \citenamefont {McMorrow}, \citenamefont {Zeiske},\
  and\ \citenamefont {Godfrey}}]{harris1997geometrical}%
  \BibitemOpen
  \bibfield  {author} {\bibinfo {author} {\bibfnamefont {M.~J.}\ \bibnamefont
  {Harris}}, \bibinfo {author} {\bibfnamefont {S.~T.}\ \bibnamefont
  {Bramwell}}, \bibinfo {author} {\bibfnamefont {D.~F.}\ \bibnamefont
  {McMorrow}}, \bibinfo {author} {\bibfnamefont {T.~H.}\ \bibnamefont
  {Zeiske}}, \ and\ \bibinfo {author} {\bibfnamefont {K.~W.}\ \bibnamefont
  {Godfrey}},\ }\href@noop {} {\bibfield  {journal} {\bibinfo  {journal} {Phys.
  Rev. Lett.}\ }\textbf {\bibinfo {volume} {79}},\ \bibinfo {pages} {2554}
  (\bibinfo {year} {1997})}\BibitemShut {NoStop}%
\bibitem [{\citenamefont {Ramirez}\ \emph {et~al.}(1999)\citenamefont
  {Ramirez}, \citenamefont {Hayashi}, \citenamefont {Cava}, \citenamefont
  {Siddharthan},\ and\ \citenamefont {Shastry}}]{ramirez1999zero}%
  \BibitemOpen
  \bibfield  {author} {\bibinfo {author} {\bibfnamefont {A.~P.}\ \bibnamefont
  {Ramirez}}, \bibinfo {author} {\bibfnamefont {A.}~\bibnamefont {Hayashi}},
  \bibinfo {author} {\bibfnamefont {R.~J.}\ \bibnamefont {Cava}}, \bibinfo
  {author} {\bibfnamefont {R.}~\bibnamefont {Siddharthan}}, \ and\ \bibinfo
  {author} {\bibfnamefont {B.~S.}\ \bibnamefont {Shastry}},\ }\href@noop {}
  {\bibfield  {journal} {\bibinfo  {journal} {Nature}\ }\textbf {\bibinfo
  {volume} {399}},\ \bibinfo {pages} {333} (\bibinfo {year}
  {1999})}\BibitemShut {NoStop}%
\bibitem [{\citenamefont {Bramwell}\ and\ \citenamefont
  {Gingras}(2001)}]{bramwell2001spin}%
  \BibitemOpen
  \bibfield  {author} {\bibinfo {author} {\bibfnamefont {S.~T.}\ \bibnamefont
  {Bramwell}}\ and\ \bibinfo {author} {\bibfnamefont {M.~J.~P.}\ \bibnamefont
  {Gingras}},\ }\href@noop {} {\bibfield  {journal} {\bibinfo  {journal}
  {Science}\ }\textbf {\bibinfo {volume} {294}},\ \bibinfo {pages} {1495}
  (\bibinfo {year} {2001})}\BibitemShut {NoStop}%
\bibitem [{\citenamefont {Castelnovo}\ \emph {et~al.}(2008)\citenamefont
  {Castelnovo}, \citenamefont {Moessner},\ and\ \citenamefont
  {Sondhi}}]{Castelnovo2008}%
  \BibitemOpen
  \bibfield  {author} {\bibinfo {author} {\bibfnamefont {C.}~\bibnamefont
  {Castelnovo}}, \bibinfo {author} {\bibfnamefont {R.}~\bibnamefont
  {Moessner}}, \ and\ \bibinfo {author} {\bibfnamefont {S.~L.}\ \bibnamefont
  {Sondhi}},\ }\href@noop {} {\bibfield  {journal} {\bibinfo  {journal}
  {Nature}\ }\textbf {\bibinfo {volume} {451}},\ \bibinfo {pages} {42}
  (\bibinfo {year} {2008})}\BibitemShut {NoStop}%
\bibitem [{\citenamefont {Morris}\ \emph {et~al.}(2009)\citenamefont {Morris},
  \citenamefont {Tennant}, \citenamefont {Grigera}, \citenamefont {Klemke},
  \citenamefont {Castelnovo}, \citenamefont {Moessner}, \citenamefont
  {Czternasty}, \citenamefont {Meissner}, \citenamefont {Rule}, \citenamefont
  {Hoffmann}, \citenamefont {Kiefer}, \citenamefont {Gerischer}, \citenamefont
  {Slobinsky},\ and\ \citenamefont {Perry}}]{morris2009dirac}%
  \BibitemOpen
  \bibfield  {author} {\bibinfo {author} {\bibfnamefont {D.~J.~P.}\
  \bibnamefont {Morris}}, \bibinfo {author} {\bibfnamefont {D.~A.}\
  \bibnamefont {Tennant}}, \bibinfo {author} {\bibfnamefont {S.~A.}\
  \bibnamefont {Grigera}}, \bibinfo {author} {\bibfnamefont {B.}~\bibnamefont
  {Klemke}}, \bibinfo {author} {\bibfnamefont {C.}~\bibnamefont {Castelnovo}},
  \bibinfo {author} {\bibfnamefont {R.}~\bibnamefont {Moessner}}, \bibinfo
  {author} {\bibfnamefont {C.}~\bibnamefont {Czternasty}}, \bibinfo {author}
  {\bibfnamefont {M.}~\bibnamefont {Meissner}}, \bibinfo {author}
  {\bibfnamefont {K.~C.}\ \bibnamefont {Rule}}, \bibinfo {author}
  {\bibfnamefont {J.-U.}\ \bibnamefont {Hoffmann}}, \bibinfo {author}
  {\bibfnamefont {K.}~\bibnamefont {Kiefer}}, \bibinfo {author} {\bibfnamefont
  {S.}~\bibnamefont {Gerischer}}, \bibinfo {author} {\bibfnamefont
  {D.}~\bibnamefont {Slobinsky}}, \ and\ \bibinfo {author} {\bibfnamefont
  {R.~S.}\ \bibnamefont {Perry}},\ }\href@noop {} {\bibfield  {journal}
  {\bibinfo  {journal} {Science}\ }\textbf {\bibinfo {volume} {326}},\ \bibinfo
  {pages} {411} (\bibinfo {year} {2009})}\BibitemShut {NoStop}%
\bibitem [{\citenamefont {Hallas}\ \emph
  {et~al.}(2017{\natexlab{a}})\citenamefont {Hallas}, \citenamefont {Gaudet},\
  and\ \citenamefont {Gaulin}}]{RevXYHallas}%
  \BibitemOpen
  \bibfield  {author} {\bibinfo {author} {\bibfnamefont {A.~M.}\ \bibnamefont
  {Hallas}}, \bibinfo {author} {\bibfnamefont {J.}~\bibnamefont {Gaudet}}, \
  and\ \bibinfo {author} {\bibfnamefont {B.~D.}\ \bibnamefont {Gaulin}},\
  }\href@noop {} {\bibfield  {journal} {\bibinfo  {journal} {In press, Annu.
  Rev. Condens. Matter Phys.}\ } (\bibinfo {year}
  {2017}{\natexlab{a}})}\BibitemShut {NoStop}%
\bibitem [{\citenamefont {Ross}\ \emph {et~al.}(2011)\citenamefont {Ross},
  \citenamefont {Savary}, \citenamefont {Gaulin},\ and\ \citenamefont
  {Balents}}]{Ross2011}%
  \BibitemOpen
  \bibfield  {author} {\bibinfo {author} {\bibfnamefont {K.~A.}\ \bibnamefont
  {Ross}}, \bibinfo {author} {\bibfnamefont {L.}~\bibnamefont {Savary}},
  \bibinfo {author} {\bibfnamefont {B.~D.}\ \bibnamefont {Gaulin}}, \ and\
  \bibinfo {author} {\bibfnamefont {L.}~\bibnamefont {Balents}},\ }\href@noop
  {} {\bibfield  {journal} {\bibinfo  {journal} {Phys. Rev. X}\ }\textbf
  {\bibinfo {volume} {1}},\ \bibinfo {pages} {021002} (\bibinfo {year}
  {2011})}\BibitemShut {NoStop}%
\bibitem [{\citenamefont {Savary}\ \emph {et~al.}(2012)\citenamefont {Savary},
  \citenamefont {Ross}, \citenamefont {Gaulin}, \citenamefont {Ruff},\ and\
  \citenamefont {Balents}}]{Savary2012}%
  \BibitemOpen
  \bibfield  {author} {\bibinfo {author} {\bibfnamefont {L.}~\bibnamefont
  {Savary}}, \bibinfo {author} {\bibfnamefont {K.~A.}\ \bibnamefont {Ross}},
  \bibinfo {author} {\bibfnamefont {B.~D.}\ \bibnamefont {Gaulin}}, \bibinfo
  {author} {\bibfnamefont {J.~P.~C.}\ \bibnamefont {Ruff}}, \ and\ \bibinfo
  {author} {\bibfnamefont {L.}~\bibnamefont {Balents}},\ }\href {\doibase
  10.1103/PhysRevLett.109.167201} {\bibfield  {journal} {\bibinfo  {journal}
  {Phys. Rev. Lett.}\ }\textbf {\bibinfo {volume} {109}},\ \bibinfo {pages}
  {167201} (\bibinfo {year} {2012})}\BibitemShut {NoStop}%
\bibitem [{\citenamefont {Gingras}\ and\ \citenamefont
  {McClarty}(2014)}]{gingras2014quantum}%
  \BibitemOpen
  \bibfield  {author} {\bibinfo {author} {\bibfnamefont {M.~J.~P.}\
  \bibnamefont {Gingras}}\ and\ \bibinfo {author} {\bibfnamefont {P.~A.}\
  \bibnamefont {McClarty}},\ }\href@noop {} {\bibfield  {journal} {\bibinfo
  {journal} {Rep. Prog. Phys.}\ }\textbf {\bibinfo {volume} {77}},\ \bibinfo
  {pages} {056501} (\bibinfo {year} {2014})}\BibitemShut {NoStop}%
\bibitem [{\citenamefont {Champion}\ \emph {et~al.}(2003)\citenamefont
  {Champion}, \citenamefont {Harris}, \citenamefont {Holdsworth}, \citenamefont
  {Wills}, \citenamefont {Balakrishnan}, \citenamefont {Bramwell},
  \citenamefont {\ifmmode \check{C}\else \v{C}\fi{}i\ifmmode~\check{z}\else
  \v{z}\fi{}m\'ar}, \citenamefont {Fennell}, \citenamefont {Gardner},
  \citenamefont {Lago}, \citenamefont {McMorrow}, \citenamefont
  {Orend\'a\ifmmode~\check{c}\else \v{c}\fi{}}, \citenamefont
  {Orend\'a\ifmmode~\check{c}\else \v{c}\fi{}ov\'a}, \citenamefont {Paul},
  \citenamefont {Smith}, \citenamefont {Telling},\ and\ \citenamefont
  {Wildes}}]{Champion2003}%
  \BibitemOpen
  \bibfield  {author} {\bibinfo {author} {\bibfnamefont {J.~D.~M.}\
  \bibnamefont {Champion}}, \bibinfo {author} {\bibfnamefont {M.~J.}\
  \bibnamefont {Harris}}, \bibinfo {author} {\bibfnamefont {P.~C.~W.}\
  \bibnamefont {Holdsworth}}, \bibinfo {author} {\bibfnamefont {A.~S.}\
  \bibnamefont {Wills}}, \bibinfo {author} {\bibfnamefont {G.}~\bibnamefont
  {Balakrishnan}}, \bibinfo {author} {\bibfnamefont {S.~T.}\ \bibnamefont
  {Bramwell}}, \bibinfo {author} {\bibfnamefont {E.}~\bibnamefont {\ifmmode
  \check{C}\else \v{C}\fi{}i\ifmmode~\check{z}\else \v{z}\fi{}m\'ar}}, \bibinfo
  {author} {\bibfnamefont {T.}~\bibnamefont {Fennell}}, \bibinfo {author}
  {\bibfnamefont {J.~S.}\ \bibnamefont {Gardner}}, \bibinfo {author}
  {\bibfnamefont {J.}~\bibnamefont {Lago}}, \bibinfo {author} {\bibfnamefont
  {D.~F.}\ \bibnamefont {McMorrow}}, \bibinfo {author} {\bibfnamefont
  {M.}~\bibnamefont {Orend\'a\ifmmode~\check{c}\else \v{c}\fi{}}}, \bibinfo
  {author} {\bibfnamefont {A.}~\bibnamefont {Orend\'a\ifmmode~\check{c}\else
  \v{c}\fi{}ov\'a}}, \bibinfo {author} {\bibfnamefont {D.~M.}\ \bibnamefont
  {Paul}}, \bibinfo {author} {\bibfnamefont {R.~I.}\ \bibnamefont {Smith}},
  \bibinfo {author} {\bibfnamefont {M.~T.~F.}\ \bibnamefont {Telling}}, \ and\
  \bibinfo {author} {\bibfnamefont {A.}~\bibnamefont {Wildes}},\ }\href
  {\doibase 10.1103/PhysRevB.68.020401} {\bibfield  {journal} {\bibinfo
  {journal} {Phys. Rev. B}\ }\textbf {\bibinfo {volume} {68}},\ \bibinfo
  {pages} {020401} (\bibinfo {year} {2003})}\BibitemShut {NoStop}%
\bibitem [{\citenamefont {Zhitomirsky}\ \emph {et~al.}(2012)\citenamefont
  {Zhitomirsky}, \citenamefont {Gvozdikova}, \citenamefont {Holdsworth},\ and\
  \citenamefont {Moessner}}]{Zhitomirsky2012}%
  \BibitemOpen
  \bibfield  {author} {\bibinfo {author} {\bibfnamefont {M.~E.}\ \bibnamefont
  {Zhitomirsky}}, \bibinfo {author} {\bibfnamefont {M.~V.}\ \bibnamefont
  {Gvozdikova}}, \bibinfo {author} {\bibfnamefont {P.~C.~W.}\ \bibnamefont
  {Holdsworth}}, \ and\ \bibinfo {author} {\bibfnamefont {R.}~\bibnamefont
  {Moessner}},\ }\href {\doibase 10.1103/PhysRevLett.109.077204} {\bibfield
  {journal} {\bibinfo  {journal} {Phys. Rev. Lett.}\ }\textbf {\bibinfo
  {volume} {109}},\ \bibinfo {pages} {077204} (\bibinfo {year}
  {2012})}\BibitemShut {NoStop}%
\bibitem [{\citenamefont {Jaubert}\ \emph {et~al.}(2015)\citenamefont
  {Jaubert}, \citenamefont {Benton}, \citenamefont {Rau}, \citenamefont
  {Oitmaa}, \citenamefont {Singh}, \citenamefont {Shannon},\ and\ \citenamefont
  {Gingras}}]{Jaubert2015}%
  \BibitemOpen
  \bibfield  {author} {\bibinfo {author} {\bibfnamefont {L.}~\bibnamefont
  {Jaubert}}, \bibinfo {author} {\bibfnamefont {O.}~\bibnamefont {Benton}},
  \bibinfo {author} {\bibfnamefont {J.~G.}\ \bibnamefont {Rau}}, \bibinfo
  {author} {\bibfnamefont {J.}~\bibnamefont {Oitmaa}}, \bibinfo {author}
  {\bibfnamefont {R.}~\bibnamefont {Singh}}, \bibinfo {author} {\bibfnamefont
  {N.}~\bibnamefont {Shannon}}, \ and\ \bibinfo {author} {\bibfnamefont
  {M.~J.~P.}\ \bibnamefont {Gingras}},\ }\href@noop {} {\bibfield  {journal}
  {\bibinfo  {journal} {Phys. Rev. Lett.}\ }\textbf {\bibinfo {volume} {115}},\
  \bibinfo {pages} {267208} (\bibinfo {year} {2015})}\BibitemShut {NoStop}%
\bibitem [{\citenamefont {Yan}\ \emph {et~al.}(2017)\citenamefont {Yan},
  \citenamefont {Benton}, \citenamefont {Jaubert},\ and\ \citenamefont
  {Shannon}}]{Yan2017}%
  \BibitemOpen
  \bibfield  {author} {\bibinfo {author} {\bibfnamefont {H.}~\bibnamefont
  {Yan}}, \bibinfo {author} {\bibfnamefont {O.}~\bibnamefont {Benton}},
  \bibinfo {author} {\bibfnamefont {L.}~\bibnamefont {Jaubert}}, \ and\
  \bibinfo {author} {\bibfnamefont {N.}~\bibnamefont {Shannon}},\ }\href@noop
  {} {\bibfield  {journal} {\bibinfo  {journal} {Phys. Rev. B}\ }\textbf
  {\bibinfo {volume} {95}},\ \bibinfo {pages} {094422} (\bibinfo {year}
  {2017})}\BibitemShut {NoStop}%
\bibitem [{\citenamefont {Dun}\ \emph {et~al.}(2015)\citenamefont {Dun},
  \citenamefont {Li}, \citenamefont {Freitas}, \citenamefont {Arrighi},
  \citenamefont {Cruz}, \citenamefont {Lee}, \citenamefont {Choi},
  \citenamefont {Cao}, \citenamefont {Silverstein}, \citenamefont {Wiebe},
  \citenamefont {Cheng},\ and\ \citenamefont {Zhou}}]{Dun2015}%
  \BibitemOpen
  \bibfield  {author} {\bibinfo {author} {\bibfnamefont {Z.~L.}\ \bibnamefont
  {Dun}}, \bibinfo {author} {\bibfnamefont {X.}~\bibnamefont {Li}}, \bibinfo
  {author} {\bibfnamefont {R.~S.}\ \bibnamefont {Freitas}}, \bibinfo {author}
  {\bibfnamefont {E.}~\bibnamefont {Arrighi}}, \bibinfo {author} {\bibfnamefont
  {C.~R.~D.}\ \bibnamefont {Cruz}}, \bibinfo {author} {\bibfnamefont
  {M.}~\bibnamefont {Lee}}, \bibinfo {author} {\bibfnamefont {E.~S.}\
  \bibnamefont {Choi}}, \bibinfo {author} {\bibfnamefont {H.~B.}\ \bibnamefont
  {Cao}}, \bibinfo {author} {\bibfnamefont {H.~J.}\ \bibnamefont
  {Silverstein}}, \bibinfo {author} {\bibfnamefont {C.~R.}\ \bibnamefont
  {Wiebe}}, \bibinfo {author} {\bibfnamefont {J.~G.}\ \bibnamefont {Cheng}}, \
  and\ \bibinfo {author} {\bibfnamefont {H.~D.}\ \bibnamefont {Zhou}},\
  }\href@noop {} {\bibfield  {journal} {\bibinfo  {journal} {Phys. Rev. B}\
  }\textbf {\bibinfo {volume} {92}},\ \bibinfo {pages} {140407} (\bibinfo
  {year} {2015})}\BibitemShut {NoStop}%
\bibitem [{\citenamefont {Poole}\ \emph {et~al.}(2007)\citenamefont {Poole},
  \citenamefont {Wills},\ and\ \citenamefont {Lelievre-Berna}}]{Poole2007}%
  \BibitemOpen
  \bibfield  {author} {\bibinfo {author} {\bibfnamefont {A.}~\bibnamefont
  {Poole}}, \bibinfo {author} {\bibfnamefont {A.~S.}\ \bibnamefont {Wills}}, \
  and\ \bibinfo {author} {\bibfnamefont {E.}~\bibnamefont {Lelievre-Berna}},\
  }\href@noop {} {\bibfield  {journal} {\bibinfo  {journal} {J. Phys. Condens.
  Matter}\ }\textbf {\bibinfo {volume} {19}},\ \bibinfo {pages} {452201}
  (\bibinfo {year} {2007})}\BibitemShut {NoStop}%
\bibitem [{\citenamefont {Hallas}\ \emph
  {et~al.}(2017{\natexlab{b}})\citenamefont {Hallas}, \citenamefont {Gaudet},
  \citenamefont {Butch}, \citenamefont {Xu}, \citenamefont {Tachibana},
  \citenamefont {Wiebe}, \citenamefont {Luke},\ and\ \citenamefont
  {Gaulin}}]{Hallas2017EPO}%
  \BibitemOpen
  \bibfield  {author} {\bibinfo {author} {\bibfnamefont {A.~M.}\ \bibnamefont
  {Hallas}}, \bibinfo {author} {\bibfnamefont {J.}~\bibnamefont {Gaudet}},
  \bibinfo {author} {\bibfnamefont {N.~P.}\ \bibnamefont {Butch}}, \bibinfo
  {author} {\bibfnamefont {G.}~\bibnamefont {Xu}}, \bibinfo {author}
  {\bibfnamefont {M.}~\bibnamefont {Tachibana}}, \bibinfo {author}
  {\bibfnamefont {C.~R.}\ \bibnamefont {Wiebe}}, \bibinfo {author}
  {\bibfnamefont {G.~M.}\ \bibnamefont {Luke}}, \ and\ \bibinfo {author}
  {\bibfnamefont {B.~D.}\ \bibnamefont {Gaulin}},\ }\href@noop {} {\bibfield
  {journal} {\bibinfo  {journal} {arXiv preprint arXiv:1705.06680}\ } (\bibinfo
  {year} {2017}{\natexlab{b}})}\BibitemShut {NoStop}%
\bibitem [{\citenamefont {Petit}\ \emph {et~al.}(2017)\citenamefont {Petit},
  \citenamefont {Lhotel}, \citenamefont {Damay}, \citenamefont {Boutrouille},
  \citenamefont {Forget},\ and\ \citenamefont {Colson}}]{Petit2017}%
  \BibitemOpen
  \bibfield  {author} {\bibinfo {author} {\bibfnamefont {S.}~\bibnamefont
  {Petit}}, \bibinfo {author} {\bibfnamefont {E.}~\bibnamefont {Lhotel}},
  \bibinfo {author} {\bibfnamefont {F.}~\bibnamefont {Damay}}, \bibinfo
  {author} {\bibfnamefont {P.}~\bibnamefont {Boutrouille}}, \bibinfo {author}
  {\bibfnamefont {A.}~\bibnamefont {Forget}}, \ and\ \bibinfo {author}
  {\bibfnamefont {D.}~\bibnamefont {Colson}},\ }\href@noop {} {\bibfield
  {journal} {\bibinfo  {journal} {arXiv preprint arXiv:1705.04462}\ } (\bibinfo
  {year} {2017})}\BibitemShut {NoStop}%
\bibitem [{\citenamefont {Hutchings}(1964)}]{Hutchings}%
  \BibitemOpen
  \bibfield  {author} {\bibinfo {author} {\bibfnamefont {M.}~\bibnamefont
  {Hutchings}},\ }in\ \href {\doibase
  http://dx.doi.org/10.1016/S0081-1947(08)60517-2} {\emph {\bibinfo {booktitle}
  {Solid State Physics}}},\ Vol.~\bibinfo {volume} {16},\ \bibinfo {editor}
  {edited by\ \bibinfo {editor} {\bibfnamefont {F.}~\bibnamefont {Seitz}}\ and\
  \bibinfo {editor} {\bibfnamefont {D.}~\bibnamefont {Turnbull}}}\ (\bibinfo
  {publisher} {Academic Press},\ \bibinfo {year} {1964})\ pp.\ \bibinfo {pages}
  {227 -- 273}\BibitemShut {NoStop}%
\bibitem [{\citenamefont {Prather}(1961)}]{Prather}%
  \BibitemOpen
  \bibfield  {author} {\bibinfo {author} {\bibfnamefont {J.}~\bibnamefont
  {Prather}},\ }\href@noop {} {\bibfield  {journal} {\bibinfo  {journal} {NBS
  monograph 19}\ } (\bibinfo {year} {1961})}\BibitemShut {NoStop}%
\bibitem [{\citenamefont {Freeman}\ and\ \citenamefont
  {Watson}(1962)}]{Freeman1962}%
  \BibitemOpen
  \bibfield  {author} {\bibinfo {author} {\bibfnamefont {A.~J.}\ \bibnamefont
  {Freeman}}\ and\ \bibinfo {author} {\bibfnamefont {R.~E.}\ \bibnamefont
  {Watson}},\ }\href {\doibase 10.1103/PhysRev.127.2058} {\bibfield  {journal}
  {\bibinfo  {journal} {Phys. Rev.}\ }\textbf {\bibinfo {volume} {127}},\
  \bibinfo {pages} {2058} (\bibinfo {year} {1962})}\BibitemShut {NoStop}%
\bibitem [{\citenamefont {Walter}(1984)}]{Walter1984}%
  \BibitemOpen
  \bibfield  {author} {\bibinfo {author} {\bibfnamefont {U.}~\bibnamefont
  {Walter}},\ }\href {\doibase http://dx.doi.org/10.1016/0022-3697(84)90147-1}
  {\bibfield  {journal} {\bibinfo  {journal} {J. Phys. Chem. Solids.}\ }\textbf
  {\bibinfo {volume} {45}},\ \bibinfo {pages} {401 } (\bibinfo {year}
  {1984})}\BibitemShut {NoStop}%
\bibitem [{\citenamefont {Stevens}(1952)}]{Stevens}%
  \BibitemOpen
  \bibfield  {author} {\bibinfo {author} {\bibfnamefont {K.~W.~H.}\
  \bibnamefont {Stevens}},\ }\href@noop {} {\bibfield  {journal} {\bibinfo
  {journal} {Proceedings of the Physical Society. Section A}\ }\textbf
  {\bibinfo {volume} {65}},\ \bibinfo {pages} {209} (\bibinfo {year}
  {1952})}\BibitemShut {NoStop}%
\bibitem [{\citenamefont {Freeman}\ and\ \citenamefont
  {Desclaux}(1979)}]{freeman1979dirac}%
  \BibitemOpen
  \bibfield  {author} {\bibinfo {author} {\bibfnamefont {A.}~\bibnamefont
  {Freeman}}\ and\ \bibinfo {author} {\bibfnamefont {J.}~\bibnamefont
  {Desclaux}},\ }\href@noop {} {\bibfield  {journal} {\bibinfo  {journal} {J.
  Magn. Magn. Mater.}\ }\textbf {\bibinfo {volume} {12}},\ \bibinfo {pages}
  {11} (\bibinfo {year} {1979})}\BibitemShut {NoStop}%
\bibitem [{\citenamefont {Squires}(1978)}]{Squires}%
  \BibitemOpen
  \bibfield  {author} {\bibinfo {author} {\bibfnamefont {G.}~\bibnamefont
  {Squires}},\ }\href@noop {} {\bibfield  {journal} {\bibinfo  {journal}
  {Cambridge University Press, Cambridge, UK}\ } (\bibinfo {year}
  {1978})}\BibitemShut {NoStop}%
\bibitem [{\citenamefont {Bertin}\ \emph {et~al.}(2012)\citenamefont {Bertin},
  \citenamefont {Chapuis}, \citenamefont {de~R{\'e}otier},\ and\ \citenamefont
  {Yaouanc}}]{bertin2012crystal}%
  \BibitemOpen
  \bibfield  {author} {\bibinfo {author} {\bibfnamefont {A.}~\bibnamefont
  {Bertin}}, \bibinfo {author} {\bibfnamefont {Y.}~\bibnamefont {Chapuis}},
  \bibinfo {author} {\bibfnamefont {P.~D.}\ \bibnamefont {de~R{\'e}otier}}, \
  and\ \bibinfo {author} {\bibfnamefont {A.}~\bibnamefont {Yaouanc}},\
  }\href@noop {} {\bibfield  {journal} {\bibinfo  {journal} {J. Phys. Condens.
  Matter}\ }\textbf {\bibinfo {volume} {24}},\ \bibinfo {pages} {256003}
  (\bibinfo {year} {2012})}\BibitemShut {NoStop}%
\bibitem [{\citenamefont {Subramanian}\ \emph {et~al.}(1983)\citenamefont
  {Subramanian}, \citenamefont {Aravamudan},\ and\ \citenamefont
  {Rao}}]{subramanian1983oxide}%
  \BibitemOpen
  \bibfield  {author} {\bibinfo {author} {\bibfnamefont {M.~A.}\ \bibnamefont
  {Subramanian}}, \bibinfo {author} {\bibfnamefont {G.}~\bibnamefont
  {Aravamudan}}, \ and\ \bibinfo {author} {\bibfnamefont {G.~V.~S.}\
  \bibnamefont {Rao}},\ }\href@noop {} {\bibfield  {journal} {\bibinfo
  {journal} {Prog. Solid. State. Ch.}\ }\textbf {\bibinfo {volume} {15}},\
  \bibinfo {pages} {55} (\bibinfo {year} {1983})}\BibitemShut {NoStop}%
\bibitem [{\citenamefont {Hallas}\ \emph
  {et~al.}(2016{\natexlab{a}})\citenamefont {Hallas}, \citenamefont {Gaudet},
  \citenamefont {Wilson}, \citenamefont {Munsie}, \citenamefont {Aczel},
  \citenamefont {Stone}, \citenamefont {Freitas}, \citenamefont
  {Arevalo-Lopez}, \citenamefont {Attfield}, \citenamefont {Tachibana},
  \citenamefont {Wiebe}, \citenamefont {Luke},\ and\ \citenamefont
  {Gaulin}}]{hallas2016xy}%
  \BibitemOpen
  \bibfield  {author} {\bibinfo {author} {\bibfnamefont {A.~M.}\ \bibnamefont
  {Hallas}}, \bibinfo {author} {\bibfnamefont {J.}~\bibnamefont {Gaudet}},
  \bibinfo {author} {\bibfnamefont {M.~N.}\ \bibnamefont {Wilson}}, \bibinfo
  {author} {\bibfnamefont {T.~J.}\ \bibnamefont {Munsie}}, \bibinfo {author}
  {\bibfnamefont {A.~A.}\ \bibnamefont {Aczel}}, \bibinfo {author}
  {\bibfnamefont {M.~B.}\ \bibnamefont {Stone}}, \bibinfo {author}
  {\bibfnamefont {R.~S.}\ \bibnamefont {Freitas}}, \bibinfo {author}
  {\bibfnamefont {A.~M.}\ \bibnamefont {Arevalo-Lopez}}, \bibinfo {author}
  {\bibfnamefont {J.~P.}\ \bibnamefont {Attfield}}, \bibinfo {author}
  {\bibfnamefont {M.}~\bibnamefont {Tachibana}}, \bibinfo {author}
  {\bibfnamefont {C.~R.}\ \bibnamefont {Wiebe}}, \bibinfo {author}
  {\bibfnamefont {G.~M.}\ \bibnamefont {Luke}}, \ and\ \bibinfo {author}
  {\bibfnamefont {B.~D.}\ \bibnamefont {Gaulin}},\ }\href@noop {} {\bibfield
  {journal} {\bibinfo  {journal} {Phys. Rev. B}\ }\textbf {\bibinfo {volume}
  {93}},\ \bibinfo {pages} {104405} (\bibinfo {year}
  {2016}{\natexlab{a}})}\BibitemShut {NoStop}%
\bibitem [{\citenamefont {Hallas}\ \emph
  {et~al.}(2016{\natexlab{b}})\citenamefont {Hallas}, \citenamefont {Sharma},
  \citenamefont {Cai}, \citenamefont {Munsie}, \citenamefont {Wilson},
  \citenamefont {Tachibana}, \citenamefont {Wiebe},\ and\ \citenamefont
  {Luke}}]{hallas2016relief}%
  \BibitemOpen
  \bibfield  {author} {\bibinfo {author} {\bibfnamefont {A.~M.}\ \bibnamefont
  {Hallas}}, \bibinfo {author} {\bibfnamefont {A.~Z.}\ \bibnamefont {Sharma}},
  \bibinfo {author} {\bibfnamefont {Y.}~\bibnamefont {Cai}}, \bibinfo {author}
  {\bibfnamefont {T.~J.}\ \bibnamefont {Munsie}}, \bibinfo {author}
  {\bibfnamefont {M.~N.}\ \bibnamefont {Wilson}}, \bibinfo {author}
  {\bibfnamefont {M.}~\bibnamefont {Tachibana}}, \bibinfo {author}
  {\bibfnamefont {C.~R.}\ \bibnamefont {Wiebe}}, \ and\ \bibinfo {author}
  {\bibfnamefont {G.~M.}\ \bibnamefont {Luke}},\ }\href@noop {} {\bibfield
  {journal} {\bibinfo  {journal} {Phys. Rev. B}\ }\textbf {\bibinfo {volume}
  {94}},\ \bibinfo {pages} {134417} (\bibinfo {year}
  {2016}{\natexlab{b}})}\BibitemShut {NoStop}%
\bibitem [{\citenamefont {Huq}\ \emph {et~al.}(2011)\citenamefont {Huq},
  \citenamefont {Hodges}, \citenamefont {Gourdon},\ and\ \citenamefont
  {Heroux}}]{huq2011powgen}%
  \BibitemOpen
  \bibfield  {author} {\bibinfo {author} {\bibfnamefont {A.}~\bibnamefont
  {Huq}}, \bibinfo {author} {\bibfnamefont {J.~P.}\ \bibnamefont {Hodges}},
  \bibinfo {author} {\bibfnamefont {O.}~\bibnamefont {Gourdon}}, \ and\
  \bibinfo {author} {\bibfnamefont {L.}~\bibnamefont {Heroux}},\ }\href@noop {}
  {\bibfield  {journal} {\bibinfo  {journal} {Zeitsch. Kristall. Proc}\
  }\textbf {\bibinfo {volume} {1}},\ \bibinfo {pages} {127} (\bibinfo {year}
  {2011})}\BibitemShut {NoStop}%
\bibitem [{\citenamefont {Rodriguez-Carvajal}(1993)}]{rodriguez1993recent}%
  \BibitemOpen
  \bibfield  {author} {\bibinfo {author} {\bibfnamefont {J.}~\bibnamefont
  {Rodriguez-Carvajal}},\ }\href@noop {} {\bibfield  {journal} {\bibinfo
  {journal} {Physica B: Condensed Matter}\ }\textbf {\bibinfo {volume} {192}},\
  \bibinfo {pages} {55} (\bibinfo {year} {1993})}\BibitemShut {NoStop}%
\bibitem [{\citenamefont {Granroth}\ \emph {et~al.}(2010)\citenamefont
  {Granroth}, \citenamefont {Kolesnikov}, \citenamefont {Sherline},
  \citenamefont {Clancy}, \citenamefont {Ross}, \citenamefont {Ruff},
  \citenamefont {Gaulin},\ and\ \citenamefont {Nagler}}]{granroth2010sequoia}%
  \BibitemOpen
  \bibfield  {author} {\bibinfo {author} {\bibfnamefont {G.~E.}\ \bibnamefont
  {Granroth}}, \bibinfo {author} {\bibfnamefont {A.~I.}\ \bibnamefont
  {Kolesnikov}}, \bibinfo {author} {\bibfnamefont {T.~E.}\ \bibnamefont
  {Sherline}}, \bibinfo {author} {\bibfnamefont {J.~P.}\ \bibnamefont
  {Clancy}}, \bibinfo {author} {\bibfnamefont {K.~A.}\ \bibnamefont {Ross}},
  \bibinfo {author} {\bibfnamefont {J.~P.~C.}\ \bibnamefont {Ruff}}, \bibinfo
  {author} {\bibfnamefont {B.~D.}\ \bibnamefont {Gaulin}}, \ and\ \bibinfo
  {author} {\bibfnamefont {S.~E.}\ \bibnamefont {Nagler}},\ }in\ \href@noop {}
  {\emph {\bibinfo {booktitle} {J. Phys. Conf. Ser.}}},\ Vol.\ \bibinfo
  {volume} {251}\ (\bibinfo {organization} {IOP Publishing},\ \bibinfo {year}
  {2010})\ p.\ \bibinfo {pages} {012058}\BibitemShut {NoStop}%
\bibitem [{\citenamefont {Arnold}\ \emph {et~al.}(2014)\citenamefont {Arnold},
  \citenamefont {Bilheux}, \citenamefont {Borreguero}, \citenamefont {Buts},
  \citenamefont {Campbell}, \citenamefont {Chapon}, \citenamefont {Doucet},
  \citenamefont {Draper}, \citenamefont {Leal}, \citenamefont {Gigg},
  \citenamefont {Lynch}, \citenamefont {Markvardsen}, \citenamefont
  {Mikkelson}, \citenamefont {Mikkelson}, \citenamefont {Miller}, \citenamefont
  {Palmen}, \citenamefont {Parker}, \citenamefont {Passos}, \citenamefont
  {Perring}, \citenamefont {Peterson}, \citenamefont {Ren}, \citenamefont
  {Reuter}, \citenamefont {Savici}, \citenamefont {Taylor}, \citenamefont
  {Taylor}, \citenamefont {Tolchenov}, \citenamefont {Zhou},\ and\
  \citenamefont {Zikovsky}}]{Mantid}%
  \BibitemOpen
  \bibfield  {author} {\bibinfo {author} {\bibfnamefont {O.}~\bibnamefont
  {Arnold}}, \bibinfo {author} {\bibfnamefont {J.}~\bibnamefont {Bilheux}},
  \bibinfo {author} {\bibfnamefont {J.}~\bibnamefont {Borreguero}}, \bibinfo
  {author} {\bibfnamefont {A.}~\bibnamefont {Buts}}, \bibinfo {author}
  {\bibfnamefont {S.}~\bibnamefont {Campbell}}, \bibinfo {author}
  {\bibfnamefont {L.}~\bibnamefont {Chapon}}, \bibinfo {author} {\bibfnamefont
  {M.}~\bibnamefont {Doucet}}, \bibinfo {author} {\bibfnamefont
  {N.}~\bibnamefont {Draper}}, \bibinfo {author} {\bibfnamefont
  {R.}~\bibnamefont {Leal}}, \bibinfo {author} {\bibfnamefont {M.}~\bibnamefont
  {Gigg}}, \bibinfo {author} {\bibfnamefont {V.}~\bibnamefont {Lynch}},
  \bibinfo {author} {\bibfnamefont {A.}~\bibnamefont {Markvardsen}}, \bibinfo
  {author} {\bibfnamefont {D.}~\bibnamefont {Mikkelson}}, \bibinfo {author}
  {\bibfnamefont {R.}~\bibnamefont {Mikkelson}}, \bibinfo {author}
  {\bibfnamefont {R.}~\bibnamefont {Miller}}, \bibinfo {author} {\bibfnamefont
  {K.}~\bibnamefont {Palmen}}, \bibinfo {author} {\bibfnamefont
  {P.}~\bibnamefont {Parker}}, \bibinfo {author} {\bibfnamefont
  {G.}~\bibnamefont {Passos}}, \bibinfo {author} {\bibfnamefont
  {T.}~\bibnamefont {Perring}}, \bibinfo {author} {\bibfnamefont
  {P.}~\bibnamefont {Peterson}}, \bibinfo {author} {\bibfnamefont
  {S.}~\bibnamefont {Ren}}, \bibinfo {author} {\bibfnamefont {M.}~\bibnamefont
  {Reuter}}, \bibinfo {author} {\bibfnamefont {A.}~\bibnamefont {Savici}},
  \bibinfo {author} {\bibfnamefont {J.}~\bibnamefont {Taylor}}, \bibinfo
  {author} {\bibfnamefont {R.}~\bibnamefont {Taylor}}, \bibinfo {author}
  {\bibfnamefont {R.}~\bibnamefont {Tolchenov}}, \bibinfo {author}
  {\bibfnamefont {W.}~\bibnamefont {Zhou}}, \ and\ \bibinfo {author}
  {\bibfnamefont {J.}~\bibnamefont {Zikovsky}},\ }\href@noop {} {\bibfield
  {journal} {\bibinfo  {journal} {Nuclear Instruments and Methods in Physics
  Research Section A: Accelerators, Spectrometers, Detectors and Associated
  Equipment}\ }\textbf {\bibinfo {volume} {764}},\ \bibinfo {pages} {156}
  (\bibinfo {year} {2014})}\BibitemShut {NoStop}%
\bibitem [{\citenamefont {Azuah}\ \emph {et~al.}(2009)\citenamefont {Azuah},
  \citenamefont {Kneller}, \citenamefont {Qiu}, \citenamefont {Brown},
  \citenamefont {Copley}, \citenamefont {Dimeo},\ and\ \citenamefont
  {Tregenna-Piggott}}]{Dave}%
  \BibitemOpen
  \bibfield  {author} {\bibinfo {author} {\bibfnamefont {R.}~\bibnamefont
  {Azuah}}, \bibinfo {author} {\bibfnamefont {L.}~\bibnamefont {Kneller}},
  \bibinfo {author} {\bibfnamefont {Y.}~\bibnamefont {Qiu}}, \bibinfo {author}
  {\bibfnamefont {C.}~\bibnamefont {Brown}}, \bibinfo {author} {\bibfnamefont
  {J.}~\bibnamefont {Copley}}, \bibinfo {author} {\bibfnamefont
  {R.}~\bibnamefont {Dimeo}}, \ and\ \bibinfo {author} {\bibfnamefont
  {P.}~\bibnamefont {Tregenna-Piggott}},\ }\href@noop {} {\bibfield  {journal}
  {\bibinfo  {journal} {J. Res. Natl. Inst. Stan. Technol.}\ }\textbf {\bibinfo
  {volume} {114}} (\bibinfo {year} {2009})}\BibitemShut {NoStop}%
\bibitem [{\citenamefont {Guitteny}\ \emph {et~al.}(2013)\citenamefont
  {Guitteny}, \citenamefont {Petit}, \citenamefont {Lhotel}, \citenamefont
  {Robert}, \citenamefont {Bonville}, \citenamefont {Forget},\ and\
  \citenamefont {Mirebeau}}]{GuittenyErSnO}%
  \BibitemOpen
  \bibfield  {author} {\bibinfo {author} {\bibfnamefont {S.}~\bibnamefont
  {Guitteny}}, \bibinfo {author} {\bibfnamefont {S.}~\bibnamefont {Petit}},
  \bibinfo {author} {\bibfnamefont {E.}~\bibnamefont {Lhotel}}, \bibinfo
  {author} {\bibfnamefont {J.}~\bibnamefont {Robert}}, \bibinfo {author}
  {\bibfnamefont {P.}~\bibnamefont {Bonville}}, \bibinfo {author}
  {\bibfnamefont {A.}~\bibnamefont {Forget}}, \ and\ \bibinfo {author}
  {\bibfnamefont {I.}~\bibnamefont {Mirebeau}},\ }\href@noop {} {\bibfield
  {journal} {\bibinfo  {journal} {Phys. Rev. B}\ }\textbf {\bibinfo {volume}
  {88}},\ \bibinfo {pages} {134408} (\bibinfo {year} {2013})}\BibitemShut
  {NoStop}%
\bibitem [{\citenamefont {Gaudet}\ \emph {et~al.}(2015)\citenamefont {Gaudet},
  \citenamefont {Maharaj}, \citenamefont {Sala}, \citenamefont {Kermarrec},
  \citenamefont {Ross}, \citenamefont {Dabkowska}, \citenamefont {Kolesnikov},
  \citenamefont {Granroth},\ and\ \citenamefont {Gaulin}}]{gaudet2015neutron}%
  \BibitemOpen
  \bibfield  {author} {\bibinfo {author} {\bibfnamefont {J.}~\bibnamefont
  {Gaudet}}, \bibinfo {author} {\bibfnamefont {D.~D.}\ \bibnamefont {Maharaj}},
  \bibinfo {author} {\bibfnamefont {G.}~\bibnamefont {Sala}}, \bibinfo {author}
  {\bibfnamefont {E.}~\bibnamefont {Kermarrec}}, \bibinfo {author}
  {\bibfnamefont {K.~A.}\ \bibnamefont {Ross}}, \bibinfo {author}
  {\bibfnamefont {H.~A.}\ \bibnamefont {Dabkowska}}, \bibinfo {author}
  {\bibfnamefont {A.~I.}\ \bibnamefont {Kolesnikov}}, \bibinfo {author}
  {\bibfnamefont {G.~E.}\ \bibnamefont {Granroth}}, \ and\ \bibinfo {author}
  {\bibfnamefont {B.~D.}\ \bibnamefont {Gaulin}},\ }\href@noop {} {\bibfield
  {journal} {\bibinfo  {journal} {Phys. Rev. B}\ }\textbf {\bibinfo {volume}
  {92}},\ \bibinfo {pages} {134420} (\bibinfo {year} {2015})}\BibitemShut
  {NoStop}%
\bibitem [{\citenamefont {Wong}\ \emph {et~al.}(2013)\citenamefont {Wong},
  \citenamefont {Hao},\ and\ \citenamefont {Gingras}}]{wong2013ground}%
  \BibitemOpen
  \bibfield  {author} {\bibinfo {author} {\bibfnamefont {A.~W.~C.}\
  \bibnamefont {Wong}}, \bibinfo {author} {\bibfnamefont {Z.}~\bibnamefont
  {Hao}}, \ and\ \bibinfo {author} {\bibfnamefont {M.~J.~P.}\ \bibnamefont
  {Gingras}},\ }\href@noop {} {\bibfield  {journal} {\bibinfo  {journal} {Phys.
  Rev. B}\ }\textbf {\bibinfo {volume} {88}},\ \bibinfo {pages} {144402}
  (\bibinfo {year} {2013})}\BibitemShut {NoStop}%
\bibitem [{\citenamefont {Rosenkranz}\ \emph {et~al.}(2000)\citenamefont
  {Rosenkranz}, \citenamefont {Ramirez}, \citenamefont {Hayashi}, \citenamefont
  {Cava}, \citenamefont {Siddharthan},\ and\ \citenamefont
  {Shastry}}]{rosenkranz2000crystal}%
  \BibitemOpen
  \bibfield  {author} {\bibinfo {author} {\bibfnamefont {S.}~\bibnamefont
  {Rosenkranz}}, \bibinfo {author} {\bibfnamefont {A.~P.}\ \bibnamefont
  {Ramirez}}, \bibinfo {author} {\bibfnamefont {A.}~\bibnamefont {Hayashi}},
  \bibinfo {author} {\bibfnamefont {R.~J.}\ \bibnamefont {Cava}}, \bibinfo
  {author} {\bibfnamefont {R.}~\bibnamefont {Siddharthan}}, \ and\ \bibinfo
  {author} {\bibfnamefont {B.~S.}\ \bibnamefont {Shastry}},\ }\href@noop {}
  {\bibfield  {journal} {\bibinfo  {journal} {J. Appl. Phys.}\ }\textbf
  {\bibinfo {volume} {87}},\ \bibinfo {pages} {5914} (\bibinfo {year}
  {2000})}\BibitemShut {NoStop}%
\bibitem [{\citenamefont {Ruminy}\ \emph {et~al.}(2016)\citenamefont {Ruminy},
  \citenamefont {Pomjakushina}, \citenamefont {Iida}, \citenamefont {Kamazawa},
  \citenamefont {Adroja}, \citenamefont {Stuhr},\ and\ \citenamefont
  {Fennell}}]{ruminy2016crystal}%
  \BibitemOpen
  \bibfield  {author} {\bibinfo {author} {\bibfnamefont {M.}~\bibnamefont
  {Ruminy}}, \bibinfo {author} {\bibfnamefont {E.}~\bibnamefont
  {Pomjakushina}}, \bibinfo {author} {\bibfnamefont {K.}~\bibnamefont {Iida}},
  \bibinfo {author} {\bibfnamefont {K.}~\bibnamefont {Kamazawa}}, \bibinfo
  {author} {\bibfnamefont {D.~T.}\ \bibnamefont {Adroja}}, \bibinfo {author}
  {\bibfnamefont {U.}~\bibnamefont {Stuhr}}, \ and\ \bibinfo {author}
  {\bibfnamefont {T.}~\bibnamefont {Fennell}},\ }\href@noop {} {\bibfield
  {journal} {\bibinfo  {journal} {Phys. Rev. B}\ }\textbf {\bibinfo {volume}
  {94}},\ \bibinfo {pages} {024430} (\bibinfo {year} {2016})}\BibitemShut
  {NoStop}%
\bibitem [{\citenamefont {Li}\ \emph {et~al.}(2013)\citenamefont {Li},
  \citenamefont {Xu}, \citenamefont {Fan}, \citenamefont {Zhang}, \citenamefont
  {Lv}, \citenamefont {Ni}, \citenamefont {Zhao},\ and\ \citenamefont
  {Sun}}]{li2013single}%
  \BibitemOpen
  \bibfield  {author} {\bibinfo {author} {\bibfnamefont {Q.}~\bibnamefont
  {Li}}, \bibinfo {author} {\bibfnamefont {L.}~\bibnamefont {Xu}}, \bibinfo
  {author} {\bibfnamefont {C.}~\bibnamefont {Fan}}, \bibinfo {author}
  {\bibfnamefont {F.}~\bibnamefont {Zhang}}, \bibinfo {author} {\bibfnamefont
  {Y.}~\bibnamefont {Lv}}, \bibinfo {author} {\bibfnamefont {B.}~\bibnamefont
  {Ni}}, \bibinfo {author} {\bibfnamefont {Z.}~\bibnamefont {Zhao}}, \ and\
  \bibinfo {author} {\bibfnamefont {X.}~\bibnamefont {Sun}},\ }\href@noop {}
  {\bibfield  {journal} {\bibinfo  {journal} {J. Cryst. Growth}\ }\textbf
  {\bibinfo {volume} {377}},\ \bibinfo {pages} {96} (\bibinfo {year}
  {2013})}\BibitemShut {NoStop}%
\bibitem [{\citenamefont {Lau}\ \emph {et~al.}(2006)\citenamefont {Lau},
  \citenamefont {Muegge}, \citenamefont {McQueen}, \citenamefont {Duncan},\
  and\ \citenamefont {Cava}}]{lau2006stuffed}%
  \BibitemOpen
  \bibfield  {author} {\bibinfo {author} {\bibfnamefont {G.}~\bibnamefont
  {Lau}}, \bibinfo {author} {\bibfnamefont {B.}~\bibnamefont {Muegge}},
  \bibinfo {author} {\bibfnamefont {T.}~\bibnamefont {McQueen}}, \bibinfo
  {author} {\bibfnamefont {E.}~\bibnamefont {Duncan}}, \ and\ \bibinfo {author}
  {\bibfnamefont {R.}~\bibnamefont {Cava}},\ }\href@noop {} {\bibfield
  {journal} {\bibinfo  {journal} {J. Solid State Chem.}\ }\textbf {\bibinfo
  {volume} {179}},\ \bibinfo {pages} {3126} (\bibinfo {year}
  {2006})}\BibitemShut {NoStop}%
\bibitem [{\citenamefont {Zhang}\ \emph {et~al.}(2014)\citenamefont {Zhang},
  \citenamefont {Fritsch}, \citenamefont {Hao}, \citenamefont {Bagheri},
  \citenamefont {Gingras}, \citenamefont {Granroth}, \citenamefont
  {Jiramongkolchai}, \citenamefont {Cava},\ and\ \citenamefont
  {Gaulin}}]{zhang2014neutron}%
  \BibitemOpen
  \bibfield  {author} {\bibinfo {author} {\bibfnamefont {J.}~\bibnamefont
  {Zhang}}, \bibinfo {author} {\bibfnamefont {K.}~\bibnamefont {Fritsch}},
  \bibinfo {author} {\bibfnamefont {Z.}~\bibnamefont {Hao}}, \bibinfo {author}
  {\bibfnamefont {B.~V.}\ \bibnamefont {Bagheri}}, \bibinfo {author}
  {\bibfnamefont {M.~J.~P.}\ \bibnamefont {Gingras}}, \bibinfo {author}
  {\bibfnamefont {G.~E.}\ \bibnamefont {Granroth}}, \bibinfo {author}
  {\bibfnamefont {P.}~\bibnamefont {Jiramongkolchai}}, \bibinfo {author}
  {\bibfnamefont {R.~J.}\ \bibnamefont {Cava}}, \ and\ \bibinfo {author}
  {\bibfnamefont {B.~D.}\ \bibnamefont {Gaulin}},\ }\href@noop {} {\bibfield
  {journal} {\bibinfo  {journal} {Phys. Rev. B}\ }\textbf {\bibinfo {volume}
  {89}},\ \bibinfo {pages} {134410} (\bibinfo {year} {2014})}\BibitemShut
  {NoStop}%
\bibitem [{\citenamefont {Princep}\ \emph {et~al.}(2015)\citenamefont
  {Princep}, \citenamefont {Walker}, \citenamefont {Adroja}, \citenamefont
  {Prabhakaran},\ and\ \citenamefont {Boothroyd}}]{princep2015crystal}%
  \BibitemOpen
  \bibfield  {author} {\bibinfo {author} {\bibfnamefont {A.~J.}\ \bibnamefont
  {Princep}}, \bibinfo {author} {\bibfnamefont {H.~C.}\ \bibnamefont {Walker}},
  \bibinfo {author} {\bibfnamefont {D.~T.}\ \bibnamefont {Adroja}}, \bibinfo
  {author} {\bibfnamefont {D.}~\bibnamefont {Prabhakaran}}, \ and\ \bibinfo
  {author} {\bibfnamefont {A.~T.}\ \bibnamefont {Boothroyd}},\ }\href@noop {}
  {\bibfield  {journal} {\bibinfo  {journal} {Phys. Rev. B}\ }\textbf {\bibinfo
  {volume} {91}},\ \bibinfo {pages} {224430} (\bibinfo {year}
  {2015})}\BibitemShut {NoStop}%
\bibitem [{\citenamefont {Gardner}\ \emph {et~al.}(1999)\citenamefont
  {Gardner}, \citenamefont {Dunsiger}, \citenamefont {Gaulin}, \citenamefont
  {Gingras}, \citenamefont {Greedan}, \citenamefont {Kiefl}, \citenamefont
  {Lumsden}, \citenamefont {MacFarlane}, \citenamefont {Raju}, \citenamefont
  {Sonier}, \citenamefont {Swainson},\ and\ \citenamefont
  {Tun}}]{gardner1999cooperative}%
  \BibitemOpen
  \bibfield  {author} {\bibinfo {author} {\bibfnamefont {J.~S.}\ \bibnamefont
  {Gardner}}, \bibinfo {author} {\bibfnamefont {S.~R.}\ \bibnamefont
  {Dunsiger}}, \bibinfo {author} {\bibfnamefont {B.~D.}\ \bibnamefont
  {Gaulin}}, \bibinfo {author} {\bibfnamefont {M.~J.~P.}\ \bibnamefont
  {Gingras}}, \bibinfo {author} {\bibfnamefont {J.~E.}\ \bibnamefont
  {Greedan}}, \bibinfo {author} {\bibfnamefont {R.~F.}\ \bibnamefont {Kiefl}},
  \bibinfo {author} {\bibfnamefont {M.~D.}\ \bibnamefont {Lumsden}}, \bibinfo
  {author} {\bibfnamefont {W.~A.}\ \bibnamefont {MacFarlane}}, \bibinfo
  {author} {\bibfnamefont {N.~P.}\ \bibnamefont {Raju}}, \bibinfo {author}
  {\bibfnamefont {J.~E.}\ \bibnamefont {Sonier}}, \bibinfo {author}
  {\bibfnamefont {I.}~\bibnamefont {Swainson}}, \ and\ \bibinfo {author}
  {\bibfnamefont {Z.}~\bibnamefont {Tun}},\ }\href@noop {} {\bibfield
  {journal} {\bibinfo  {journal} {Phys. Rev. Lett.}\ }\textbf {\bibinfo
  {volume} {82}},\ \bibinfo {pages} {1012} (\bibinfo {year}
  {1999})}\BibitemShut {NoStop}%
\bibitem [{\citenamefont {Gingras}\ \emph {et~al.}(2000)\citenamefont
  {Gingras}, \citenamefont {Den~Hertog}, \citenamefont {Faucher}, \citenamefont
  {Gardner}, \citenamefont {Dunsiger}, \citenamefont {Chang}, \citenamefont
  {Gaulin}, \citenamefont {Raju},\ and\ \citenamefont
  {Greedan}}]{gingras2000thermodynamic}%
  \BibitemOpen
  \bibfield  {author} {\bibinfo {author} {\bibfnamefont {M.~J.~P.}\
  \bibnamefont {Gingras}}, \bibinfo {author} {\bibfnamefont {B.~C.}\
  \bibnamefont {Den~Hertog}}, \bibinfo {author} {\bibfnamefont
  {M.}~\bibnamefont {Faucher}}, \bibinfo {author} {\bibfnamefont {J.~S.}\
  \bibnamefont {Gardner}}, \bibinfo {author} {\bibfnamefont {S.~R.}\
  \bibnamefont {Dunsiger}}, \bibinfo {author} {\bibfnamefont {L.~J.}\
  \bibnamefont {Chang}}, \bibinfo {author} {\bibfnamefont {B.~D.}\ \bibnamefont
  {Gaulin}}, \bibinfo {author} {\bibfnamefont {N.~P.}\ \bibnamefont {Raju}}, \
  and\ \bibinfo {author} {\bibfnamefont {J.~E.}\ \bibnamefont {Greedan}},\
  }\href@noop {} {\bibfield  {journal} {\bibinfo  {journal} {Phys. Rev. B}\
  }\textbf {\bibinfo {volume} {62}},\ \bibinfo {pages} {6496} (\bibinfo {year}
  {2000})}\BibitemShut {NoStop}%
\bibitem [{\citenamefont {Molavian}\ \emph {et~al.}(2007)\citenamefont
  {Molavian}, \citenamefont {Gingras},\ and\ \citenamefont
  {Canals}}]{molavian2007dynamically}%
  \BibitemOpen
  \bibfield  {author} {\bibinfo {author} {\bibfnamefont {H.~R.}\ \bibnamefont
  {Molavian}}, \bibinfo {author} {\bibfnamefont {M.~J.~P.}\ \bibnamefont
  {Gingras}}, \ and\ \bibinfo {author} {\bibfnamefont {B.}~\bibnamefont
  {Canals}},\ }\href@noop {} {\bibfield  {journal} {\bibinfo  {journal} {Phys.
  Rev. Lett.}\ }\textbf {\bibinfo {volume} {98}},\ \bibinfo {pages} {157204}
  (\bibinfo {year} {2007})}\BibitemShut {NoStop}%
\bibitem [{\citenamefont {Rau}\ \emph {et~al.}(2016)\citenamefont {Rau},
  \citenamefont {Petit},\ and\ \citenamefont {Gingras}}]{rau2016order}%
  \BibitemOpen
  \bibfield  {author} {\bibinfo {author} {\bibfnamefont {J.~G.}\ \bibnamefont
  {Rau}}, \bibinfo {author} {\bibfnamefont {S.}~\bibnamefont {Petit}}, \ and\
  \bibinfo {author} {\bibfnamefont {M.~J.~P.}\ \bibnamefont {Gingras}},\
  }\href@noop {} {\bibfield  {journal} {\bibinfo  {journal} {Phys. Rev. B}\
  }\textbf {\bibinfo {volume} {93}},\ \bibinfo {pages} {184408} (\bibinfo
  {year} {2016})}\BibitemShut {NoStop}%
\bibitem [{\citenamefont {McClarty}\ \emph {et~al.}(2009)\citenamefont
  {McClarty}, \citenamefont {Curnoe},\ and\ \citenamefont
  {Gingras}}]{mcclarty2009energetic}%
  \BibitemOpen
  \bibfield  {author} {\bibinfo {author} {\bibfnamefont {P.~A.}\ \bibnamefont
  {McClarty}}, \bibinfo {author} {\bibfnamefont {S.~H.}\ \bibnamefont
  {Curnoe}}, \ and\ \bibinfo {author} {\bibfnamefont {M.~J.~P.}\ \bibnamefont
  {Gingras}},\ }in\ \href@noop {} {\emph {\bibinfo {booktitle} {J. Phys. Conf.
  Ser.}}},\ Vol.\ \bibinfo {volume} {145}\ (\bibinfo {organization} {IOP
  Publishing},\ \bibinfo {year} {2009})\ p.\ \bibinfo {pages}
  {012032}\BibitemShut {NoStop}%
\bibitem [{\citenamefont {Petit}\ \emph {et~al.}(2014)\citenamefont {Petit},
  \citenamefont {Robert}, \citenamefont {Guitteny}, \citenamefont {Bonville},
  \citenamefont {Decorse}, \citenamefont {Ollivier}, \citenamefont {Mutka},
  \citenamefont {Gingras},\ and\ \citenamefont {Mirebeau}}]{petit2014order}%
  \BibitemOpen
  \bibfield  {author} {\bibinfo {author} {\bibfnamefont {S.}~\bibnamefont
  {Petit}}, \bibinfo {author} {\bibfnamefont {J.}~\bibnamefont {Robert}},
  \bibinfo {author} {\bibfnamefont {S.}~\bibnamefont {Guitteny}}, \bibinfo
  {author} {\bibfnamefont {P.}~\bibnamefont {Bonville}}, \bibinfo {author}
  {\bibfnamefont {C.}~\bibnamefont {Decorse}}, \bibinfo {author} {\bibfnamefont
  {J.}~\bibnamefont {Ollivier}}, \bibinfo {author} {\bibfnamefont
  {H.}~\bibnamefont {Mutka}}, \bibinfo {author} {\bibfnamefont {M.~J.~P.}\
  \bibnamefont {Gingras}}, \ and\ \bibinfo {author} {\bibfnamefont
  {I.}~\bibnamefont {Mirebeau}},\ }\href@noop {} {\bibfield  {journal}
  {\bibinfo  {journal} {Phys. Rev. B}\ }\textbf {\bibinfo {volume} {90}},\
  \bibinfo {pages} {060410} (\bibinfo {year} {2014})}\BibitemShut {NoStop}%
\end{thebibliography}%

\begin{table*}[]
\centering
\caption{Tables giving the experimental and calculated values of the seven CEF excited states energies relative to the CEF ground state as well as the relative scattered intensities. The relative scattered intensity has been normalized by the intensity of the transition between the CEF ground state and the first excited state.}
\label{Exp_value}
\begin{tabular}{lccccllcccc}
\cline{1-5} \cline{7-11}
\multicolumn{1}{|l}{}   & \multicolumn{4}{c|}{Er$_2$Ge$_2$O$_7$}                                                                                                                        & \multicolumn{1}{l|}{} &                        & \multicolumn{4}{c|}{Er$_2$Ti$_2$O$_7$}                                                                                                                        \\ \cline{1-5} \cline{7-11} 
\multicolumn{1}{|l|}{}  & \multicolumn{1}{c|}{$E_{exp}$ (meV)} & \multicolumn{1}{c|}{$E_{calc}$ (meV)} & \multicolumn{1}{c|}{$I_{exp}$ (a.u.)} & \multicolumn{1}{c|}{$I_{calc}$ (a.u.)} & \multicolumn{1}{l|}{} & \multicolumn{1}{c|}{}  & \multicolumn{1}{c|}{$E_{exp}$ (meV)} & \multicolumn{1}{c|}{$E_{calc}$ (meV)} & \multicolumn{1}{c|}{$I_{exp}$ (a.u.)} & \multicolumn{1}{c|}{$I_{calc}$ (a.u.)}  \\ \cline{1-5} \cline{7-11} 
\multicolumn{1}{|l|}{1} & \multicolumn{1}{c|}{6.6(1)}          & \multicolumn{1}{c|}{6.5}              & \multicolumn{1}{c|}{1}                & \multicolumn{1}{c|}{1}                 & \multicolumn{1}{l|}{} & \multicolumn{1}{l|}{1} & \multicolumn{1}{c|}{6.3(1)}          & \multicolumn{1}{c|}{6.4}              & \multicolumn{1}{c|}{1}                & \multicolumn{1}{c|}{1}                 \\ \cline{1-5} \cline{7-11} 
\multicolumn{1}{|l|}{2} & \multicolumn{1}{c|}{9.3(1)}          & \multicolumn{1}{c|}{9.1}              & \multicolumn{1}{c|}{0.65(5)}          & \multicolumn{1}{c|}{0.57}              & \multicolumn{1}{l|}{} & \multicolumn{1}{l|}{2} & \multicolumn{1}{c|}{7.3(1)}          & \multicolumn{1}{c|}{7.3}              & \multicolumn{1}{c|}{0.75(5)}          & \multicolumn{1}{c|}{0.88}              \\ \cline{1-5} \cline{7-11} 
\multicolumn{1}{|l|}{3} & \multicolumn{1}{c|}{20.2(1)}         & \multicolumn{1}{c|}{20.3}             & \multicolumn{1}{c|}{0.20(3)}          & \multicolumn{1}{c|}{0.24}              & \multicolumn{1}{l|}{} & \multicolumn{1}{l|}{3} & \multicolumn{1}{c|}{15.7(1)}         & \multicolumn{1}{c|}{15.7}             & \multicolumn{1}{c|}{0.20(5)}          & \multicolumn{1}{c|}{0.3}               \\ \cline{1-5} \cline{7-11} 
\multicolumn{1}{|l|}{4} & \multicolumn{1}{c|}{69.2(3)}         & \multicolumn{1}{c|}{69.2}             & \multicolumn{1}{c|}{0.04(3)}          & \multicolumn{1}{c|}{0.04}              & \multicolumn{1}{l|}{} & \multicolumn{1}{l|}{4} & \multicolumn{1}{c|}{60.2(3)}         & \multicolumn{1}{c|}{60.6}             & \multicolumn{1}{c|}{0.04(2)}          & \multicolumn{1}{c|}{0.07}              \\ \cline{1-5} \cline{7-11} 
\multicolumn{1}{|l|}{5} & \multicolumn{1}{c|}{71.1(3)}         & \multicolumn{1}{c|}{70.8}             & \multicolumn{1}{c|}{0.04(3)}          & \multicolumn{1}{c|}{0.04}              & \multicolumn{1}{l|}{} & \multicolumn{1}{l|}{5} & \multicolumn{1}{c|}{62.3(3)}         & \multicolumn{1}{c|}{62.0}             & \multicolumn{1}{c|}{0.04(2)}          & \multicolumn{1}{c|}{0.05}              \\ \cline{1-5} \cline{7-11} 
\multicolumn{1}{|l|}{6} & \multicolumn{1}{c|}{75.8(3)}         & \multicolumn{1}{c|}{75.6}             & \multicolumn{1}{c|}{0.10(2)}          & \multicolumn{1}{c|}{0.14}              & \multicolumn{1}{l|}{} & \multicolumn{1}{l|}{6} & \multicolumn{1}{c|}{66.3(3)}         & \multicolumn{1}{c|}{65.9}             & \multicolumn{1}{c|}{0.09(2)}          & \multicolumn{1}{c|}{0.13}              \\ \cline{1-5} \cline{7-11} 
\multicolumn{1}{|l|}{7} & \multicolumn{1}{c|}{95.3(4)}         & \multicolumn{1}{c|}{95.3}             & \multicolumn{1}{c|}{0.015(5)}         & \multicolumn{1}{c|}{0.03}              & \multicolumn{1}{l|}{} & \multicolumn{1}{l|}{7} & \multicolumn{1}{c|}{87.2(4)}         & \multicolumn{1}{c|}{86.7}             & \multicolumn{1}{c|}{0.01(1)}          & \multicolumn{1}{c|}{0.04}              \\ \cline{1-5} \cline{7-11} 
                        & \multicolumn{1}{l}{}                 & \multicolumn{1}{l}{}                  & \multicolumn{1}{l}{}                  & \multicolumn{1}{l}{}                   &                       &                        & \multicolumn{1}{l}{}                 & \multicolumn{1}{l}{}                  & \multicolumn{1}{l}{}                  & \multicolumn{1}{l}{}                   \\ \cline{1-5} \cline{7-11} 
\multicolumn{1}{|l}{}   & \multicolumn{4}{c|}{Er$_2$Pt$_2$O$_7$}                                                                                                                        & \multicolumn{1}{l|}{} &                        & \multicolumn{4}{c|}{Er$_2$Sn$_2$O$_7$}                                                                                                                        \\ \cline{1-5} \cline{7-11} 
\multicolumn{1}{|l|}{}  & \multicolumn{1}{c|}{$E_{exp}$ (meV)} & \multicolumn{1}{c|}{$E_{calc}$ (meV)} & \multicolumn{1}{c|}{$I_{exp}$ (a.u.)} & \multicolumn{1}{c|}{$I_{calc}$ (a.u.)} & \multicolumn{1}{l|}{} & \multicolumn{1}{c|}{}  & \multicolumn{1}{c|}{$E_{exp}$ (meV)} & \multicolumn{1}{c|}{$E_{calc}$ (meV)} & \multicolumn{1}{c|}{$I_{exp}$ (a.u.)} & \multicolumn{1}{c|}{$I_{calc}$ (a.u.)} \\ \cline{1-5} \cline{7-11} 
\multicolumn{1}{|l|}{1} & \multicolumn{1}{c|}{5.5(1)}          & \multicolumn{1}{c|}{5.7}              & \multicolumn{1}{c|}{1}                & \multicolumn{1}{c|}{1}                 & \multicolumn{1}{l|}{} & \multicolumn{1}{l|}{1} & \multicolumn{1}{c|}{5.0}             & \multicolumn{1}{c|}{5.0}              & \multicolumn{1}{c|}{1}                & \multicolumn{1}{c|}{1}                 \\ \cline{1-5} \cline{7-11} 
\multicolumn{1}{|l|}{2} & \multicolumn{1}{c|}{9.5(1)}          & \multicolumn{1}{c|}{9.1}              & \multicolumn{1}{c|}{0.29(5)}          & \multicolumn{1}{c|}{0.33}              & \multicolumn{1}{l|}{} & \multicolumn{1}{l|}{2} & \multicolumn{1}{c|}{7.4(1)}          & \multicolumn{1}{c|}{7.3}              & \multicolumn{1}{c|}{0.36(5)}          & \multicolumn{1}{c|}{0.43}              \\ \cline{1-5} \cline{7-11} 
\multicolumn{1}{|l|}{3} & \multicolumn{1}{c|}{21.2(1)}         & \multicolumn{1}{c|}{21.3}             & \multicolumn{1}{c|}{0.21(5)}          & \multicolumn{1}{c|}{0.28}              & \multicolumn{1}{l|}{} & \multicolumn{1}{l|}{3} & \multicolumn{1}{c|}{17.3(1)}         & \multicolumn{1}{c|}{17.5}             & \multicolumn{1}{c|}{0.24(5)}          & \multicolumn{1}{c|}{0.3}               \\ \cline{1-5} \cline{7-11} 
\multicolumn{1}{|l|}{4} & \multicolumn{1}{c|}{64.1(3)}         & \multicolumn{1}{c|}{63.7}             & \multicolumn{1}{c|}{0.04(2)}          & \multicolumn{1}{c|}{0.03}              & \multicolumn{1}{l|}{} & \multicolumn{1}{l|}{4} & \multicolumn{1}{c|}{55.9(3)}         & \multicolumn{1}{c|}{55.8}             & \multicolumn{1}{c|}{0.04(2)}          & \multicolumn{1}{c|}{0.04}              \\ \cline{1-5} \cline{7-11} 
\multicolumn{1}{|l|}{5} & \multicolumn{1}{c|}{65.6(3)}         & \multicolumn{1}{c|}{65.1}             & \multicolumn{1}{c|}{0.06(2)}          & \multicolumn{1}{c|}{0.06}              & \multicolumn{1}{l|}{} & \multicolumn{1}{l|}{5} & \multicolumn{1}{c|}{57.9(3)}         & \multicolumn{1}{c|}{57.5}             & \multicolumn{1}{c|}{0.06(2)}          & \multicolumn{1}{c|}{0.07}              \\ \cline{1-5} \cline{7-11} 
\multicolumn{1}{|l|}{6} & \multicolumn{1}{c|}{68.8(3)}         & \multicolumn{1}{c|}{69.5}             & \multicolumn{1}{c|}{0.12(2)}          & \multicolumn{1}{c|}{0.16}              & \multicolumn{1}{l|}{} & \multicolumn{1}{l|}{6} & \multicolumn{1}{c|}{66.8(3)}         & \multicolumn{1}{c|}{61.2}             & \multicolumn{1}{c|}{0.10(2)}          & \multicolumn{1}{c|}{0.2}               \\ \cline{1-5} \cline{7-11} 
\multicolumn{1}{|l|}{7} & \multicolumn{1}{c|}{88.7(4)}         & \multicolumn{1}{c|}{88.7}             & \multicolumn{1}{c|}{0.01(1)}          & \multicolumn{1}{c|}{0.02}              & \multicolumn{1}{l|}{} & \multicolumn{1}{l|}{7} & \multicolumn{1}{c|}{81.8(4)}         & \multicolumn{1}{c|}{81.1}             & \multicolumn{1}{c|}{0.02(1)}          & \multicolumn{1}{c|}{0.02}              \\ \cline{1-5} \cline{7-11} 
\end{tabular}
\end{table*}

\end{document}